\documentclass[aps,prd,reprint,superscriptaddress,nofootinbib]{revtex4-2}

\usepackage{amsmath,amssymb}
\usepackage{graphicx}
\usepackage{hyperref}
\usepackage[dvipsnames]{xcolor}
\usepackage{booktabs}
\usepackage{multirow}
\usepackage{siunitx}
\usepackage{orcidlink}
\usepackage{bm}

\newcommand{\Msun}{M_\odot}
\newcommand{\chieff}{\chi_\mathrm{eff}}
\newcommand{\chip}{\chi_p}
\newcommand{\nrsur}{\textsc{NRSur7dq4}}
\newcommand{\dd}{\mathrm{d}}
\newcommand{\MM}{\mathcal{M}}

\providecommand{\mnras}{Mon. Not. R. Astron. Soc.}

\begin{document}

\title{Fast, accurate, and differentiable: a neural-network surrogate for NRSur7dq4 precessing binary black hole waveforms}

\newcommand{\URI}{\affiliation{Department of Physics, East Hall,
        University of Rhode Island, Kingston, RI 02881, USA}}
\newcommand{\uriIACR}{\affiliation{Institute for AI \& Computational Research,
        Carothers Library, University of Rhode Island, Kingston, RI 02881, USA}}
\newcommand{\UMassD}{\affiliation{Department of Mathematics,
    Center for Scientific Computing and Data Science Research,
    University of Massachusetts, Dartmouth, MA 02747, USA}}

\newcommand{\TAPIR}{\affiliation{TAPIR, California Institute of Technology, Pasadena, California 91125, USA}}
\newcommand{\LIGOLab}{\affiliation{LIGO Laboratory, California Institute of Technology, Pasadena, California 91125, USA}}

\author{Michael P\"urrer~\orcidlink{0000-0002-3329-9788}}
\URI
\uriIACR

\author{Ashwin Girish~\orcidlink{0009-0004-5115-5667}}
\URI

\author{Lucy M.~Thomas~\orcidlink{0000-0003-3271-6436}}
\TAPIR
\LIGOLab

\author{Scott E. Field~\orcidlink{0000-0002-6037-3277}}
\UMassD

\author{Vijay Varma~\orcidlink{0000-0002-9994-1761}}
\UMassD

\date{\today}

\begin{abstract}
We present a neural network surrogate model that emulates the \nrsur{}
gravitational waveform model for precessing binary black hole mergers.
The surrogate decomposes the waveform into constituent quantities
and trains an independent multilayer
perceptron (MLP) for each.
We validate the surrogate against \nrsur{} on 10\,000 waveforms spanning
its full parameter space ($1 \leq q \leq 4$,
$|\boldsymbol{\chi}_{A,B}| \leq 0.8$).  For representative total masses
between $60$ and $300\,\Msun$, median sky-averaged frequency-domain
mismatches range from $8.0 \times 10^{-5}$ to $1.7 \times 10^{-4}$,
with 95th percentiles below $10^{-3}$.
On an NVIDIA L40S GPU the JAX surrogate evaluates a single waveform in
$\sim 1\,\mathrm{ms}$ end-to-end ($\sim 10\times$ faster than the 
\textsc{lalsimulation} C implementation of \nrsur{}) and sustains $\sim 140\times$
the \textsc{lalsimulation} throughput at batch size 64, making it well
suited for both low-latency parameter-estimation samplers and large-scale
waveform generation.
The full NRSur7dq4 NN waveform-to-likelihood pipeline is implemented in JAX and
is differentiable. This is the first \emph{neural-network} surrogate of
a precessing numerical-relativity waveform model to combine validated NR-faithful accuracy
with a fully differentiable, GPU-accelerated inference pipeline, enabling gradient-based
inference approaches via automatic differentiation including Fisher information matrices,
GPU-accelerated nested sampling, gradient-based MCMC and importance sampling.
\end{abstract}

\maketitle

\section{Introduction}
\label{sec:intro}

The first direct detection of gravitational waves (GWs) from a binary black hole (BBH) merger by LIGO in 2015~\cite{LIGOScientific:2016aoc} marked the beginning of the era of GW astronomy. 
Since then, the LIGO-Virgo-KAGRA (LVK) detector network~\cite{LIGOScientific:2014pky,VIRGO:2014yos,KAGRA:2019htd} has observed over 400 GW events~\cite{LIGOScientific:2026sit,LIGOScientific:2026ifv,LIGOScientific:2026wfs}, the majority of which are BBH mergers.
These events have provided unprecedented insights into the nature of black holes and the dynamics of their mergers.
The accurate modeling of the GW signals emitted by these mergers is crucial to locating these events and extracting 
the physical parameters of the source, such as the masses and spins of the black holes.
The most accurate models for these signals are generated by numerical relativity (NR) simulations~\cite{Scheel:2025jct,Pretorius:2005gq,Campanelli:2005dd,Baker:2005vv}, which solve the full Einstein equations numerically.
However, these simulations are expensive to run, often requiring weeks to generate a waveform for a single system. 
GW parameter estimation (PE) requires millions of waveforms to be generated, making it impractical to use NR simulations directly for inference.

Several alternative strategies have therefore been developed to model waveforms. 
Phenomenological models reduce the computational cost by fitting the amplitude and phase of
the spherical harmonic modes of the waveform to piecewise analytic functions~\cite{Garcia-Quiros:2020qpx,Ajith:2009bn,Khan:2015jqa,Pratten:2020ceb}.
The effective-one-body (EOB) approach maps the two-body problem in general relativity onto
the dynamics of a test mass moving in a deformed Kerr spacetime, calibrated to NR
simulations~\cite{Ramos-Buades:2023ehm,Buonanno:1998gg,Ossokine:2020kjp,Albanesi:2025txj}. Both strategies make physical approximations.

A complementary strategy is surrogate or reduced order modeling~\cite{Field:2011mf,Field:2013cfa,Purrer:2014fza,Purrer:2015tud,Blackman:2015pia,Blackman:2017pcm,Cotesta:2020qhw}. Surrogate models are data-driven approximations to more expensive underlying models,
constructed by sampling the model at a set of training points in the parameter space and using regression techniques to predict the model output at new parameter points.
One could build a surrogate model on top of any waveform model, including phenomenological and EOB models, to further speed up evaluation.
Building a surrogate directly on NR simulations allows one to achieve accuracy 
comparable to NR itself, at a fraction of the evaluation cost~\cite{Varma:2019csw_phm}. \nrsur{}~\cite{Varma:2019csw} 
is the state-of-the-art NR surrogate model for generic precessing BBH systems, covering 
mass ratios up to $q = 4$ and spin magnitudes up to $0.8$. It is currently one of the 
most accurate waveform models available for precessing systems and has been used extensively
in LVK PE analyses~\cite{LIGOScientific:2026ifv}.

The next generation of ground-based detectors like the Einstein Telescope~\cite{Punturo:2010zz,Branchesi:2023mws,ET:2025xjr} and Cosmic Explorer~\cite{Reitze:2019iox,Evans:2023euw} will be capable of
observing GWs from all stellar-mass BBH systems in the observable universe, dramatically increasing both the number of detected events and their signal-to-noise ratios (SNRs).
Both trends compound the computational burden of PE. The number of analyses grows with the event rate, while each individual analysis becomes more demanding because a higher SNR sharpens the posterior and requires more waveform evaluations to resolve it.
Fully exploiting the increased sensitivity of these detectors also demands more accurate waveform models~\cite{Lindblom:2008cm,Purrer:2019jcp,Hu:2022rjq,Read:2023hkv,Owen:2023mid,Kapil:2024zdn,Dhani:2024jja}, whose improved fidelity has historically come at the price of increased evaluation latency.
Since each likelihood evaluation in turn requires generating a waveform, the speed of the waveform model directly sets the achievable PE turnaround.

Several approaches have been tested to reduce the cost of likelihood evaluation, including heterodyning, also known as relative binning~\cite{Cornish:2010kf,Zackay:2018qdy,Cornish:2021lje,Leslie:2021ssu},
multi-banding ~\cite{Vinciguerra:2017ngf, Morisaki:2021ngj}, and reduced order quadratures ~\cite{Canizares:2013ywa, Canizares:2014fya, Smith:2016qas, Tissino:2022thn}. 
One could also bypass the direct evaluation of the likelihood function during the inference process,
using simulation-based inference methods~\cite{Chua:2019wwt,Dax:2021tsq,Dax:2022pxd,Dax:2024mcn,Hu:2024lrj}, which instead train neural networks on large sets of simulated waveforms.
Here the training set is generated only once and its cost amortized over all subsequent analyses; the step that is genuinely latency-bound is the importance sampling used to verify and reweight the resulting posteriors~\cite{Dax:2022pxd}, which evaluates the waveform model once per posterior sample.
Across all of these approaches, fast and accurate waveform generation remains a central requirement.
To this end, neural surrogate models such as those presented in~\cite{Khan:2020fso,Thomas:2022rmc,GramaxoFreitas:2024bpk,Thomas:2025rje}
have proven successful in accelerating waveform evaluation while preserving waveform fidelity.
These models use neural networks to learn the mapping from the binary parameters to the waveform, 
and are trained on a large set of waveforms generated using the underlying model they seek to emulate.

In this paper, we present a neural surrogate model that emulates the \nrsur{} waveform model for precessing BBH systems.
We target the full parameter space of the underlying model, \nrsur{}, which includes mass ratios up to $q = 4$ and spin magnitudes up to $0.8$.
Our model adopts the same waveform decomposition strategy as \nrsur{} itself, 
modeling the orbital frequency, co-precessing quaternions, and co-orbital modes with
independent feed-forward neural networks, and reassembling the waveform using the same 
rotation and mode-summation pipeline. By predicting directly at each time step rather
than compressing onto a reduced basis, our model resolves the low-amplitude inspiral
more accurately than a reduced-basis representation does for our orbital-frequency
target (Sec.~\ref{sec:architecture}), while delivering up to $150\times$ speedup over the original \nrsur{}
implementation for batched evaluation on GPUs. This makes it suitable for both low-latency 
parameter estimation and large-scale training-set generation for SBI pipelines.

Concurrent work by Whittall \& Pratten~\cite{Whittall:2026yvp}
develops a neural surrogate for \textsc{SEOBNRv5PHM}
covering mass ratios up to $q = 10$ with generic spin orientations and magnitudes up to the extremal limit, and a longer pre-merger waveform duration of $\sim 10^4\,M$. Their approach shares the same high-level decomposition into co-rotating-frame modes, orbital phase, and rotation quaternions, with independent feed-forward networks for each data piece.
A key methodological difference is that their approach compresses each
time series using a reduced basis and empirical interpolation~\cite{Barrault2004,Field:2013cfa}, whereas
we predict the sampled data pieces directly. The two approaches also
use different dynamical targets: they predict the orbital phase
directly, while we predict the orbital frequency and spin trajectories
and recover the phase by integration. We provide a more detailed
comparison in Sec.~\ref{sec:discussion}.
Independent of Ref.~\cite{Whittall:2026yvp}, we also note ongoing, as-yet-unpublished work on a direct JAX re-implementation of the classical \nrsur{} spline/EIM evaluation~\cite{JAXNRSur}.

This paper is structured as follows. In Sec.~\ref{sec:nrsur} we review the structure of \nrsur{}, its parameter space, evaluation cost, and establish notation. 
Sec.~\ref{sec:methods} describes our waveform decomposition, neural network architectures, training procedure, and waveform reassembly pipeline. 
In Sec.~\ref{sec:results} we report the accuracy of individual data pieces and complete waveforms, evaluation speed, and results from PE runs comparing
our surrogate against the original \nrsur{} implementation. We summarize our findings and discuss future directions in Sec.~\ref{sec:discussion}.
We use geometric units $G = 1 = c$ in this work, unless otherwise stated.

\section{The \nrsur{} waveform model}
\label{sec:nrsur}

\nrsur{}~\cite{Varma:2019csw} is a numerical-relativity surrogate model
for the gravitational waveform from precessing binary black hole (BBH)
mergers, built from 1528 NR simulations performed with the
\textsc{SpEC} code~\cite{Lovelace:2008tw, Lindblom:2005qh,
Szilagyi:2009qz,Scheel:2008rj,Scheel:2025jct}.
It covers the last $\sim 4300\,M$ before merger
(roughly 20 orbits for an equal-mass system), where $M = m_1 + m_2$ is
the total Christodoulou mass.  We briefly review its structure and establish
notation used throughout this paper.

\subsection{Parameter space}

The model is parameterized by the seven-dimensional intrinsic parameter vector
\begin{equation}
\boldsymbol{\theta}
= \bigl(q, \chi_{1x}, \chi_{1y}, \chi_{1z}, \chi_{2x}, \chi_{2y}, \chi_{2z}\bigr) ,
\label{eq:params}
\end{equation}
where $q=m_1/m_2\geq 1$ is the mass ratio, 
with $m_1\geq m_2$, and $\boldsymbol{\chi}_1$ and $\boldsymbol{\chi}_2$ are 
the dimensionless spin vectors of the heavier and 
lighter black holes, respectively. The spin components are specified 
at the reference time $t_{\rm ref}=-4300M$ and $t=0$ corresponds
to the peak of the total waveform amplitude. 
The spin vectors are measured in the inertial frame 
used by \nrsur{}: At $t_{\rm ref}$, the $\hat{z}$ direction
of the inertial frame is along the principal eigenvector
of the angular momentum operator
with the $\hat{x}$ axis pointing from the lighter to the heavier black hole,
and the $\hat{y}$ axis completing the right-handed triad~\cite{Varma:2019csw}.
The valid range of the model is $q \in [1, 4]$ and
$|\boldsymbol{\chi}_{1}|, |\boldsymbol{\chi}_{2}| \leq 0.8$.

Several derived spin quantities are useful for characterizing the
waveform morphology and the accuracy of the surrogate. The
\emph{effective aligned spin}
\begin{equation}
  \chieff = \frac{m_1 \,\chi_{1z} + m_2 \, \chi_{2z}}{M} \,,
\label{eq:chieff}
\end{equation}
is a mass-weighted projection of the spins onto the orbital angular
momentum, approximately conserved at 2PN order, which controls the
inspiral rate.
Orbital precession is characterized by,
\begin{equation}
  \chip = \max\!\Bigl(\chi_{1,\perp},\;
    \frac{4 + 3q}{q\,(4q + 3)}\,\chi_{2,\perp}\Bigr) \,,
\label{eq:chip}
\end{equation}
which depends on the in-plane spin magnitudes
$\chi_{i,\perp}=\sqrt{\chi_{ix}^2+\chi_{iy}^2}$ of both black holes.
In addition to the intrinsic parameters, the observed waveform depends
on extrinsic quantities that are not modeled by the surrogate but
applied at waveform assembly time. These include the total mass
$M = m_1 + m_2$ (in units of $M_\odot$), which sets the physical time
and frequency scale via $t_\mathrm{phys} = t\, M \,M_\odot[s]$
where $M_\odot[s] \approx 4.93\,\mu\mathrm{s}$ is the solar mass in
seconds, the inclination angle $\iota$
between the line of sight and the orbital angular momentum $\mathbf{L}$
at the reference epoch, and the reference phase $\phi_\mathrm{ref}$.

\subsection{Model structure}
\label{sec:nrsur_structure}

\nrsur{} adopts a two-step architecture.  First, a \emph{dynamics
system} produces time-dependent quantities on a co-orbital frame time
grid $\{t_k\}_{k=1}^{T}$ with $T = 230$ non-uniformly spaced
nodes~\footnote{The original \nrsur{} paper~\cite{Varma:2019csw} incorrectly states $T=238$. 
The correct number should be $T=233$, but three nodes are initial half-step nodes used for the three RK4 startup steps of the AB4
time integrator.} in dimensionless units of $M$.  These quantities are:
\begin{enumerate}
  \item The \emph{orbital phase} $\phi_\mathrm{orb}(t)$: the
        accumulated azimuthal angle of the binary orbit.
  \item The \emph{co-precessing quaternion}
        $\mathbf{q}(t) = (q_0, q_1, q_2, q_3)(t)$: a unit quaternion
        ($|\mathbf{q}| = 1$) encoding the rotation from the
        co-precessing frame (which tracks the instantaneous orbital
        plane) to the inertial frame.
  \item The \emph{spin trajectories}
        $\boldsymbol{\chi}_A^\mathrm{copr}(t)$ and
        $\boldsymbol{\chi}_B^\mathrm{copr}(t)$: three-component spin
        vectors in the co-precessing frame.
\end{enumerate}

Second, the model evaluates the \emph{co-orbital frame modes}.
The co-orbital frame is obtained from the co-precessing frame by an
additional rotation about the $z$-axis by the orbital phase
$\phi_\mathrm{orb}(t)$, giving
\begin{equation}
  h_{\ell m}^\mathrm{coorb}(t) =
    e^{-im\phi_\mathrm{orb}(t)}\, h_{\ell m}^\mathrm{copr}(t) \,.
\label{eq:coorb_mode}
\end{equation}
Stripping the rapid orbital oscillation $e^{-im\phi_\mathrm{orb}}$
leaves modes that vary slowly in time, encoding only the amplitude
and shape modulations of the waveform, which makes them well suited
for neural-network regression. This yields a total of 21 complex modes with $\ell \leq 4$.

The inertial-frame waveform is then assembled by rotating the
co-orbital modes using Wigner $D$-matrices constructed from the
quaternion and orbital phase (Sec.~\ref{sec:assembly}).  The
gravitational-wave polarizations,
\begin{equation}
  h_+(t) - i\,h_\times(t) = \sum_{\ell m}
    {}_{-2}Y_{\ell m}\!\left(\iota,\,\tfrac{\pi}{2} - \phi_\mathrm{ref}\right)
    h_{\ell m}^\mathrm{inert}(t) \,,
\label{eq:strain}
\end{equation}
are obtained by projecting onto spin-weighted spherical harmonics. 
Here $\iota$ is the inclination angle between the \emph{orbital} angular
momentum $\mathbf{L}$ at the reference epoch and the line of sight to the
observer, and $\phi_\mathrm{ref}$ is the reference azimuthal phase
following the \textsc{lal} convention~\cite{lal}, which maps to azimuthal
angle $\pi/2 - \phi_\mathrm{ref}$ in the spherical harmonic argument.

\subsection{Computational cost}

On a single CPU core (Intel Xeon 6730P Granite Rapids), evaluating \nrsur{}
at a total mass of $M=60 M_\odot$ and sampling frequency $f_s = 4096$ Hz takes
$\sim 10\,\mathrm{ms}$ via the C implementation in \textsc{lalsimulation}~\cite{lal}.
Recent optimizations~\footnote{These optimizations were carried out as part of Ref.~\cite{Ravishankar:2026_NRSur7dq4v2} 
and released as version 1.1.9 of \textsc{gwsurrogate}.}
to the Python \textsc{gwsurrogate} package~\cite{Field:2025isp}
bring its single-waveform evaluation to a comparable speed.
This timing serves as our baseline for speed comparisons in Sec.~\ref{sec:speed}.

\section{Methods}
\label{sec:methods}

\subsection{Inner products, overlap, and mismatch}
\label{sec:mismatch_defs}

We employ two sets of waveform comparison metrics throughout this work.
Given an inner product $\langle \cdot | \cdot \rangle$, the normalized
overlap between two waveforms $h_1$ and $h_2$ is
\begin{equation}
  \mathcal{O}
    = \frac{|\langle h_1 | h_2 \rangle|}
           {\sqrt{\langle h_1 | h_1 \rangle\;
                  \langle h_2 | h_2 \rangle}} \,,
\label{eq:overlap}
\end{equation}
and the mismatch is $\MM = 1 - \mathcal{O}$.

\subsubsection{Time-domain mismatch}
The flat (unweighted) time-domain inner product between two complex
waveforms $a(t)$ and $b(t)$ is
\begin{equation}
  \langle a | b \rangle_\mathrm{TD}
    = \int a^*(t)\, b(t)\, \dd t \,,
\label{eq:td_inner}
\end{equation}
approximated on the \nrsur{} sparse time grid $\{t_k\}$ via the
trapezoidal rule.  This does not require interpolation to a uniform
grid and is differentiable with respect to the data-piece outputs, making it
suitable for sensitivity analysis (Sec.~\ref{sec:error_decomp_method}).
The time-domain mismatch $\MM_\mathrm{TD}$ is obtained from
Eq.~\eqref{eq:overlap} with this inner product, maximizing
analytically over a constant phase shift $\phi_c$.

\subsubsection{Frequency-domain inner product}
The noise-weighted frequency-domain inner product is
\begin{equation}
  \langle a | b \rangle_\mathrm{FD}
    = 4\,\mathrm{Re} \int_{f_\mathrm{min}}^{f_\mathrm{max}}
      \frac{\tilde{a}^*(f)\, \tilde{b}(f)}{S_n(f)}\, \dd f \,,
\label{eq:fd_inner}
\end{equation}
where $\tilde{a}(f)$ denotes the Fourier transform and $S_n(f)$ is
the one-sided noise power spectral density (PSD). We use the aLIGO
sensitivity curve \texttt{aLIGOLateHighSensitivityP1200087}~\cite{KAGRA:2013rdx}
available through \texttt{LALSimulation}~\cite{lal}.
Following the SXS-style convention~\cite{Varma:2019csw,Blackman:2017pcm},
the integration band $[f_\mathrm{min}, f_\mathrm{max}]$ is set by the
waveform rather than by a fixed detector band: $f_\mathrm{min}$ is the
$(2,2)$ gravitational-wave frequency at the end of the roll-on taper,
and $f_\mathrm{max}$ is the frequency at peak amplitude scaled by the
largest azimuthal index ($m = 4$) so that all harmonics through
$\ell = 4$ are covered. The total mass $M$ enters only through the
conversion between dimensionless time (units of $M$) and physical
seconds, which slides the waveform's fixed dimensionless band across the
PSD.

To evaluate Eq.~\eqref{eq:fd_inner} from the \nrsur{} sparse-grid
output, we (i) interpolate via cubic spline onto a uniform grid with
step size $\Delta t = 0.1\,M$ spanning $t \in [-4000, 100]\,M$, 
(ii) apply a Planck taper~\cite{Blackman:2017pcm} that rolls on over the
first $500\,M$ and rolls off at the end of the window, $100\,M$ past the
amplitude peak,  (iii) FFT and determine
$[f_\mathrm{min}, f_\mathrm{max}]$ from the tapered reference waveform, 
and (iv) compute the discrete inner product
$\langle a | b \rangle_\mathrm{FD} = 4\,\Delta f \sum_k
\tilde{a}_k^*\,\tilde{b}_k / S_n(f_k)$ over that band, converting
dimensionless time to physical seconds via
$\Delta t_\mathrm{phys} = \Delta t \times M \, M_\odot[s]$ for a given
total mass $M$.

\subsubsection{Sky-averaged FD mismatch}
Following the procedure standard in the NR surrogate
community~\cite{Varma:2019csw,Blackman:2017pcm}, we compute an
SXS-style sky-averaged~\footnote{
  Despite the name ``sky-averaged'' here $(\theta,\varphi)$ are source-frame
  angles, namely the direction of the line of sight relative to the binary
  (the arguments of the ${}_{-2}Y_{\ell m}$, with polar angle the inclination),
  not astronomical sky coordinates (right ascension and declination). The
  average is over the orientation of the source as seen by the observer,
  independent of detector antenna response.
} noise-weighted mismatch.  At each of
$N_\mathrm{sky} = 27$ points distributed over the 2-sphere
$(\theta,\varphi)$, the complex strain is constructed from the 21
individual inertial modes via spin-weighted spherical
harmonics,
\begin{equation}
  h_+ - i\,h_\times = \sum_{\ell m}
    {}_{-2}Y_{\ell m}(\theta,\varphi)\, h^\mathrm{inert}_{\ell m} \,,
\label{eq:strain_modes}
\end{equation}
yielding the two polarizations $h_+(\theta,\varphi)$ and
$h_\times(\theta,\varphi)$.
The mismatch at each sky point is 
minimized via Nelder--Mead over the three physical
degeneracies: (i) a time shift $t_c$, maximized via IFFT,
(ii) a polarization angle $\psi$, which combines the two polarizations
as $h = h_+ \cos 2\psi + h_\times \sin 2\psi$, and (iii) an orbital
phase offset $\delta\varphi$, which
enters through the spherical harmonics as an $m$-dependent phase shift
$e^{-im\delta\varphi}$ on each mode.
The mismatch $\MM(\theta,\varphi)$ at each sky point is obtained from
Eq.~\eqref{eq:overlap} with the FD inner product~\eqref{eq:fd_inner},
maximized over $t_c$, $\psi$, and $\delta\varphi$.
The sky-averaged mismatch is the arithmetic mean over all sky points,
\begin{equation}
  \langle \MM \rangle_\mathrm{sky}
    = \frac{1}{N_\mathrm{sky}}
      \sum_{k=1}^{N_\mathrm{sky}} \MM(\theta_k,\varphi_k) \,.
\label{eq:sky_avg}
\end{equation}
We evaluate at three total masses
$M \in \{60, 120, 300\}\,\Msun$ to probe different frequency bands of
the detector sensitivity.

\subsection{Piecewise decomposition}
\label{sec:decomposition}

\begin{figure*}[t]
  \centering
  \includegraphics[width=\textwidth]{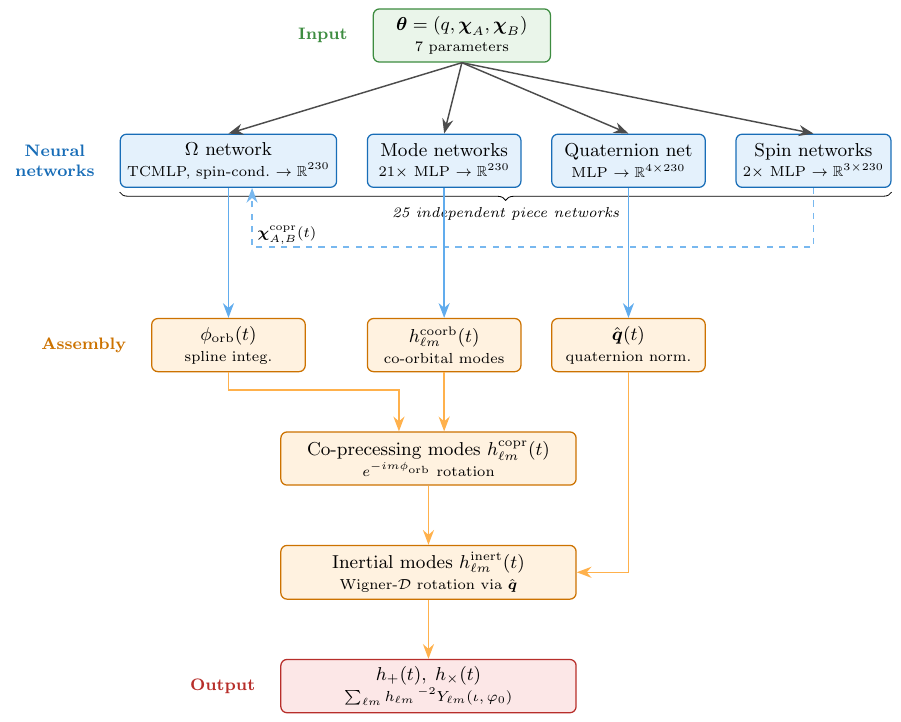}
  \caption{Overview of the piecewise neural surrogate.  The 7D binary
    parameters $\boldsymbol{\theta}$ are fed to 25 independent networks,
    one for each data piece, whose outputs are assembled into
    inertial-frame strain
    polarizations through the same rotation and mode-summation pipeline
    used by \nrsur{}.}
  \label{fig:pipeline}
\end{figure*}

Our surrogate mirrors the two-step architecture of \nrsur{}
(Sec.~\ref{sec:nrsur_structure}) by decomposing the waveform into
independent \emph{data pieces} (hereafter, \emph{pieces}), each modeled
by a separate neural network, as shown in Fig.~\ref{fig:pipeline}.
The model for piece $i$ is a function
$f_i : \boldsymbol{\theta} \in \mathbb{R}^7 \to
\mathbf{y}_i \in \mathbb{R}^{C_i \times T}$, where $C_i$ is the
number of output channels for piece $i$.
Concretely, we train $N_\mathrm{pieces} = 25$ models (cf. Fig.~\ref{fig:pipeline}) for the following pieces:
\begin{enumerate}
  \item \textbf{Orbital frequency} $\Omega(t)$: 1 channel, $T$ time
        steps.  This is the derivative of the orbital phase,
        $\Omega = \dd\phi_\mathrm{orb}/\dd t$;  we model $\Omega$
        rather than $\phi_\mathrm{orb}$ directly and recover the phase
        by numerical integration at assembly time
        (Sec.~\ref{sec:assembly}).
  \item \textbf{Co-precessing quaternion} $\mathbf{q}(t)$: 4 channels.
  \item \textbf{Spin trajectories}
        $\boldsymbol{\chi}_A^\mathrm{copr}(t)$ and
        $\boldsymbol{\chi}_B^\mathrm{copr}(t)$: 3 channels each
        (2~models).
        This differs from \nrsur{}, whose dynamics surrogate represents the
        spins in the co-orbital frame.
  \item \textbf{Co-orbital modes}~\footnote{
        We model $h^\mathrm{coorb}_{\ell m}$ directly rather than the $h^\pm_{\ell m}$
        combinations used in the original NRSur7dq4 construction. Since the mode
        networks are not the accuracy bottleneck, this simpler choice has no
         practical impact on waveform fidelity.
        }
        $h_{\ell m}^\mathrm{coorb}(t)$: 2 channels
        (real and imaginary parts) per mode, for each of the 21
        $(\ell, m)$ combinations with $\ell \leq 4$.
\end{enumerate}

This decomposition has several advantages over training a single
monolithic network for the full waveform.  First, each piece can use a
different architecture and capacity. For example, the quaternion, which enters the
Wigner $D$-rotation for all modes, requires a large network
(23\,M~parameters), while subdominant $\ell = 4$ modes are adequately
captured with 0.3\,M~parameters (Table~\ref{tab:piece_summary}).
Second, errors can be attributed to individual pieces
(Sec.~\ref{sec:error_decomp_method}), enabling targeted improvements.
Third, individual pieces can be upgraded without retraining the full
model. Fourth, modeling mostly non-oscillatory pieces is a much simpler
problem than modeling the inertial frame waveform modes directly.
Finally, unlike NRSur7dq4, which fits the instantaneous derivatives
of the orbital phase, coprecessing-frame rotation, and spins and advances
the resulting coupled system sequentially with an ODE integrator, we
predict the quaternion and spin trajectories directly from the initial
binary parameters. This removes the inherently sequential dynamics solve,
exposes the full time dimension to GPU batching, and simplifies automatic
differentiation.

\subsection{Network architecture: PieceMLP}
\label{sec:architecture}

\subsubsection{Architecture}
Our primary architecture is a standard feedforward multilayer perceptron
(MLP), which we call \emph{PieceMLP}.  Before the forward pass, the
input $\boldsymbol{\theta}$ is linearly normalized to $[-1,1]^7$.  After
the forward pass, the raw output is de-standardized using per-channel,
per-timestep statistics from the training set (details below).  
Each component of $\boldsymbol{\theta}$ is linearly mapped to $[-1, 1]$ as follows:
the mass ratio as $q_\mathrm{norm} = 2(q-1)/3 - 1$ (mapping
$[1,4] \to [-1,1]$) and the spin components as
$\chi_{i,\mathrm{norm}} = \chi_i / 0.8$ (mapping
$[-0.8, 0.8] \to [-1, 1]$).

The network maps the normalized parameter vector to all $C \times T$ output
values in a single forward pass, and naturally vectorizes over a batch
of $B$ parameter vectors via standard matrix operations:
\begin{multline}
  \boldsymbol{\theta}_\mathrm{norm}
  \;\xrightarrow{\;W_0,\,b_0\;}
  \mathrm{GELU}
  \;\xrightarrow[\times\,(D{-}1)]{\;W_i,\,b_i\;}
  \mathrm{GELU} \\
  \xrightarrow{\;W_D,\,b_D\;}
  \mathbf{z} \in \mathbb{R}^{C \times T},
\label{eq:mlp}
\end{multline}
where $W_0 \in \mathbb{R}^{H \times 7}$ is the input projection,
$W_i \in \mathbb{R}^{H \times H}$ are hidden-layer weights with width
$H$, $D$ is the number of hidden layers, and the GELU activation is
$\mathrm{GELU}(x) = x\,\Phi(x)$, where $\Phi$ is the standard normal
CDF.  This smooth, non-monotonic activation empirically outperforms ReLU
for our regression tasks.

\subsubsection{Per-timestep output standardization}
The raw network output $\mathbf{z}$ is \emph{de-standardized} to
physical units via
\begin{equation}
  y_c(t_k) = \mu_c(t_k) + \sigma_c(t_k) \cdot z_c(t_k) \,,
\label{eq:destandardize}
\end{equation}
where $\mu_c(t_k)$ and $\sigma_c(t_k)$ are the per-channel,
per-timestep mean and standard deviation computed from the training set
and stored as frozen (non-trainable) buffers.  This ensures that the
network's learning targets have approximately zero mean and unit
variance at each time step, compensating for the large dynamic range of
the physical outputs.  For example, for the smoothed orbital-frequency
target the per-timestep standard deviation $\sigma(t_k)$ varies by a
factor of $\sim 36$ across the time grid (Sec.~\ref{sec:smoothing}).

\subsubsection{Width versus depth}
The hyperparameter sweeps in Appendix~\ref{app:sweeps} 
consistently show that increasing the width $H$ at fixed 
depth $D$ improves accuracy more than increasing the depth 
at fixed width. For the quaternion model, a network with 
$H=2048$ and $D=6$ achieves a validation MSE of 
$1.04\times10^{-6}$, compared with $2.27\times10^{-6}$ 
for $H=1024$ and $D=12$. Thus, at a comparable parameter 
count, doubling the width rather than the depth improves 
the validation MSE by a factor of $2.2$. Deep networks 
with $D\geq12$ also suffer from vanishing gradients when 
residual connections are not used. A mode-training run 
with $H=512$ and $D=12$ shows no learning, while the 
corresponding $H=1024$ network begins to converge only 
after approximately 130 epochs. We therefore adopt
shallow, wide architectures with $D=4$--$6$ as the default.
The orbital-frequency TCMLP is the only deeper model in the production
bank and uses residual connections to support stable training
(Sec.~\ref{sec:tcmlp}).

\subsubsection{Comparison with a reduced-basis representation}
The original \nrsur{} surrogate~\cite{Blackman:2017pcm} constructs reduced bases and
empirical interpolants for its co-orbital waveform data pieces and fits their values at
selected empirical time nodes.  We tested an analogous representation for the
orbital-frequency target $\Omega(t)$ by building a rank-$K = 46$ SVD basis (retaining
$99.99\%$ of the training-set variance), selecting $K$ empirical nodes, and training a
PieceMLP to predict either the orthogonal-projection coefficients or the coefficients
obtained from values at the empirical nodes.  Holding the architecture, training set, and
target fixed (PieceMLP $512\times12$, $5\times10^6$ waveforms, smoothed $\Omega$), direct
per-timestep prediction attains a physical MSE\footnote{Physical MSE here is the mean
over 10\,000 evaluation waveforms of
the per-waveform mean squared error $\langle(\hat{\Omega}(t;\boldsymbol{\lambda}) -
\Omega(t;\boldsymbol{\lambda}))^2\rangle_t$ over the 230 stored time nodes, in units of
$M^{-2}$.  It differs from the dimensionless training MSE quoted elsewhere, which is
standardized by the per-timestep variance of the training targets.}
of $3.3\times10^{-8}\,M^{-2}$, compared with
$1.2\times10^{-7}\,M^{-2}$ for the SVD-coefficient model and
$1.6\times10^{-7}\,M^{-2}$ for the EIM-coefficient model.

Reconstruction from exact target data separates representation error from
coefficient-regression error.  On the validation set this gives a full-grid MSE of
$2.6\times10^{-8}\,M^{-2}$ for the rank-46 orthogonal projection and
$7.4\times10^{-8}\,M^{-2}$ for the exact empirical interpolant.

The limitations of these reduced-basis representations are most
apparent during the inspiral.
For $t \leq 0$, the exact projection and interpolation errors are
$2.6\times10^{-8}\,M^{-2}$ and $6.0\times10^{-8}\,M^{-2}$, respectively.
The projection error is essentially unchanged from its full-grid value, because a
basis built to capture total variance is dominated by the high-amplitude merger and
ringdown and leaves a nearly flat absolute error across the time grid.
In comparison, the direct model reaches $5.1\times10^{-10}\,M^{-2}$, a factor of $50$
below the exact-projection error and $120$ below the exact-interpolation error.
Because the inspiral carries most of the signal-to-noise for the sources of
interest, while ringdown errors are strongly suppressed by the detector PSD, the

unweighted full-grid MSE quoted above understates the advantage of
direct prediction during the inspiral.
Larger $K$ would lower these errors, at the cost of the compression the
representation is meant to provide.  We therefore adopt direct, per-timestep prediction
of $\Omega(t)$.  We make no claim that this ordering transfers to other data pieces or
targets, ranks, weightings, or basis constructions.

\subsection{Time-conditioned MLP for orbital frequency}
\label{sec:tcmlp}

We found the orbital frequency $\Omega(t)$
to be among the more challenging dynamical quantities to model accurately.
Its accuracy is also particularly consequential because errors in $\Omega(t)$
accumulate when it is integrated to recover the orbital phase.
We therefore explored an
alternative architecture, the \emph{time-conditioned MLP}
(TCMLP), that treats time as an explicit input rather than encoding
the full time series in a single output layer.

The distinction between the two architectures can be summarized as
\begin{align*}
  \mathrm{PieceMLP:}\;
  \boldsymbol{\theta}
  &\longmapsto
  [\hat{\Omega}(t_1),\ldots,\hat{\Omega}(t_T)], \\
  \mathrm{TCMLP:}\;
  \bigl(\boldsymbol{\theta},\mathrm{FFE}(t_k),
  \boldsymbol{\chi}_{A,B}^\mathrm{copr}(t_k)\bigr)
  &\longmapsto \hat{\Omega}(t_k).
\end{align*}
Intuitively, PieceMLP treats time as an output label: a single
evaluation maps the binary parameters to a vector whose $k$th entry
represents $\Omega(t_k)$.  TCMLP instead treats time as an input
coordinate and learns a single shared map that can be evaluated at every
time node. In the production TCMLP model, time is represented by a temporal
Fourier feature encoding (FFE) and the two spin vectors at that
time. The FFE and spin conditioning are described below.

\subsubsection{Architecture}
The TCMLP transforms the time input via a Fourier feature
encoding (FFE),
\begin{multline}
  \mathrm{FFE}(t) = \bigl(t_\mathrm{norm},\;
    \sin(2\pi\nu_1 t_\mathrm{norm}),\,
    \cos(2\pi\nu_1 t_\mathrm{norm}),\, \\
    \ldots,\,
    \sin(2\pi\nu_F t_\mathrm{norm}),\,
    \cos(2\pi\nu_F t_\mathrm{norm})
  \bigr) \,,
\label{eq:ffe}
\end{multline}
where $t_\mathrm{norm} = (t - t_\mathrm{min}) / (t_\mathrm{max} -
t_\mathrm{min}) \in [0, 1]$, and $\nu_j$ are $F = 16$ log-spaced
frequencies (in cycles per unit normalized time) ranging from $\nu_1 = 1$
to $\nu_F = T/2$.  
The FFE has dimension $1 + 2F = 33$. We denote the total network input
dimension by $D_\mathrm{in}$.
Unlike the PieceMLP, which predicts all $T$ timesteps with a large output layer,
the TCMLP model predicts $C$ scalar values per pair.
At inference, all $T$ timesteps are batched schematically as
$(B \times T,\, D_\mathrm{in}) \to \mathrm{MLP} \to (B \times T,\, C)$.

On a uniformly spaced $T$-point grid $\nu_F$ would be the
Nyquist frequency, but the \nrsur{} grid is not uniform: it is close to uniform
in orbital phase, with ${\sim}10$ nodes per orbit, so the spacing runs from
$37.6\,M$ at the start of the waveform to $2.0\,M$ near merger.  The shortest
period represented by the encoding, $(t_\mathrm{max}-t_\mathrm{min})/\nu_F
\approx 38\,M$, is therefore below the local Nyquist limit of $75\,M$ in the
early inspiral and far above the $4\,M$ limit near merger.  The highest
encoded frequencies are thus resolved only over part of the grid, which is
consistent with the sweeps of Appendix~\ref{app:sweeps}, where $F = 16$
outperforms $F = 32$.

Two separate considerations motivate this design.  The first concerns time
conditioning itself.  The PieceMLP output layer treats the $T$ time samples as
exchangeable slots, with nothing in the architecture encoding the fact that
$t_k$ and $t_{k+1}$ are adjacent. The TCMLP, instead, learns a single map that is
shared across the grid and can be evaluated at any $t$.  The second concerns the
encoding of that input.  
Standard MLPs tend to learn slowly varying dependence on their
inputs before rapid variations, a behavior known as spectral
bias~\cite{pmlr-v97-rahaman19a}. Fourier features expose the network to
several temporal scales and make rapid variation with time easier to
learn~\cite{Tancik2020}.

This is the only piece for which we depart from the shallow default of
Sec.~\ref{sec:architecture}: the production $\Omega$ model uses $D = 16$,
whereas every other piece in the bank uses $D = 4$--$6$
(Table~\ref{tab:piece_summary}).  For depths $D > 12$ we replace the plain
hidden layers with \emph{pre-activation residual blocks}.  Each block applies two
weight matrices $W_1, W_2 \in \mathbb{R}^{H \times H}$ and adds the
result back to the input:
\begin{equation}
  \mathbf{x} \;\mapsto\; \mathbf{x} +
    W_2\,\mathrm{GELU}(W_1\,\mathrm{GELU}(\mathbf{x})) \,,
\label{eq:resblock}
\end{equation}
where the skip connection $\mathbf{x} + (\cdots)$ is what makes this a
residual block.

The input projection is followed by two-layer residual blocks,
with one additional hidden layer when needed to obtain the specified
depth $D$.

Plain deep TCMLPs failed to learn in our sweeps, whereas residual
variants trained stably at depths up to $D=24$.

\subsubsection{Spin conditioning}
The TCMLP for the orbital-frequency data piece uses per-timestep spin
trajectories as additional input:
$\boldsymbol{\chi}_A^\mathrm{copr}(t_k)$ and
$\boldsymbol{\chi}_B^\mathrm{copr}(t_k)$ are concatenated to the
Fourier-feature-encoded input, adding 6 dimensions per timestep.  This
mimics the \nrsur{} ODE integrator, whose right-hand side depends on
the instantaneous spin state.  During training the ground-truth spin
trajectories are used. At inference, predictions from the pre-trained
spin models are substituted.  The total input dimension is thus
$D_\mathrm{in} = 7 + 33 + 6 = 46$.

\subsubsection{Comparison with PieceMLP}
Table~\ref{tab:tcmlp_vs_piecemlp} compares the two architectures on
the orbital-frequency data piece.  For models of comparable size trained
on 2\,M samples, the TCMLP with spin
conditioning and PieceMLP achieve nearly identical accuracy
($3.83 \times 10^{-6}$ vs $3.85 \times 10^{-6}$).  However, at
5\,M training samples with the smoothed $\Omega$ target
(Sec.~\ref{sec:smoothing}), the TCMLP outperforms PieceMLP by 37\%
($2.55 \times 10^{-5}$ vs $4.06 \times 10^{-5}$), using 25\% fewer
parameters.

Spin conditioning consistently provides a $\sim 14$--$16\%$ improvement
across TCMLP configurations (e.g., $5.29 \times 10^{-6} \to
4.56 \times 10^{-6}$ at $H = 256$, $D = 20$).

Despite its accuracy advantage, the TCMLP incurs $\sim 2\times$ higher
inference cost per piece (due to the $B \times T$ forward-pass
expansion) and introduces an inference-time dependency on pre-trained
spin models.  The PieceMLP, by contrast, predicts the entire time
series in a single forward pass, stacks naturally in the parallel model
bank (Sec.~\ref{sec:bank}), and requires no auxiliary models.  For the
production bank, we use PieceMLP for all data pieces except $\Omega$, where the
TCMLP's accuracy advantage justifies the added complexity.

\begin{table}[t]
  \centering
  \caption{PieceMLP vs TCMLP model for the orbital-frequency data piece.
    Each network's configuration is quoted as ($H \times D$), where $H$ is the hidden width and $D$ is the depth.
    The TCMLP model uses Fourier feature encoding, residual connections, and spin conditioning ( ``SC''). Val MSE is the dimensionless validation MSE on 100k held-out samples,
    i.e.\ the mean squared error of the standardized network output
    $z_c(t_k)$ in Eq.~\eqref{eq:destandardize}, equivalently the
    physical MSE divided by the per-timestep training-set variance
    $\sigma_c^2(t_k)$.  We quote the standardized MSE because an unweighted
    physical MSE over the full grid is dominated by the ringdown, where
    errors are strongly suppressed by the detector PSD.
    ``Smooth'' uses the smoothed target $\Omega_\mathrm{flt}$
    (Sec.~\ref{sec:smoothing}).}
  \label{tab:tcmlp_vs_piecemlp}
  \footnotesize
  \begin{ruledtabular}
  \begin{tabular}{lrcc}
    Model & Params
      & \multicolumn{2}{c}{Val MSE} \\
      &        & 2M, raw $\Omega$ & 5M, smooth $\Omega$ \\
    \hline
    PieceMLP ($512\times12$)     & 3.01M
      & $3.85\times10^{-6}$ & $4.06\times10^{-5}$ \\
    TCMLP\,+\,SC ($384\times16$) & 2.24M
      & $3.83\times10^{-6}$ & $2.55\times10^{-5}$ \\
  \end{tabular}
  \end{ruledtabular}
\end{table}

\subsection{Orbital phase smoothing}
\label{sec:smoothing}

We find that the quality of the \emph{training target} for the orbital
frequency is at least as important as the network architecture or
dataset size.

The naive target $\Omega_\mathrm{raw}(t) = \dd\phi_\mathrm{orb}/\dd t$,
obtained by differentiating the \nrsur{} orbital phase with a cubic
spline, is noisy in the post-merger region ($t \gtrsim 40\,M$) where the
orbital-phase concept becomes ill-defined and numerical artifacts are
amplified by differentiation.  The per-timestep standard deviation
$\sigma(t_k)$ computed across the training set varies by a factor
$\sim 210$ over the time grid, from $9.9 \times 10^{-4}$ in the early
inspiral to $0.21$ near merger.

\begin{figure}[t]
  \centering
  \includegraphics[width=\columnwidth]{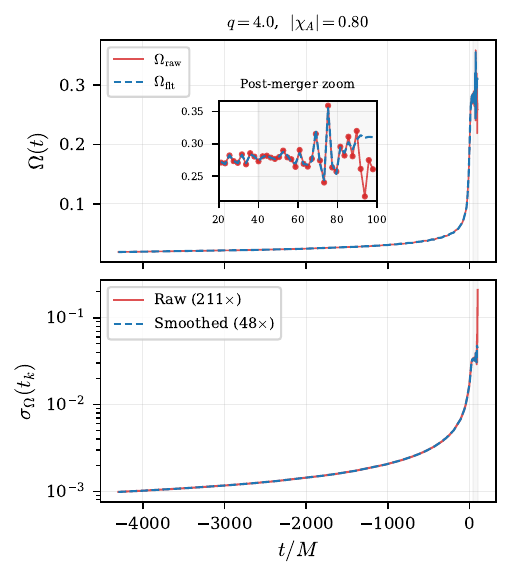}
  \caption{Effect of post-merger orbital-phase smoothing.
    \emph{Top:} Raw (dashed) and smoothed (solid) orbital frequency
    $\Omega(t)$ for a representative high-spin waveform.
    \emph{Bottom:} Per-timestep standard deviation $\sigma(t_k)$
    across the training set.  Smoothing reduces the dynamic range of
    $\sigma$ from $\sim 211\times$ to $\sim 48\times$, concentrated
    in the post-merger region $t \gtrsim 40\,M$.}
  \label{fig:smoothing}
\end{figure}

To mitigate this, we apply a post-merger smoothing procedure to the
orbital phase before differentiation:
We first upsample $\phi_\mathrm{orb}(t)$ to a $10\times$ finer grid
and compute its derivative using finite differences.  In the
post-merger window $t \in [40,90]\,M$, we replace the fine-grid
derivative with a zero-smoothing-factor ($s=0$) interpolating spline
fitted to the derivative values, and for $t>90\,M$ extend the
derivative as a constant.  Finally, we integrate this modified
derivative to obtain $\phi_\mathrm{flt}(t)$ and differentiate it on the
original sparse grid to define $\Omega_\mathrm{flt}(t)$.
This reduces the $\sigma(t_k)$ dynamic range from $\sim 210\times$ to
$\sim 48\times$ (maximum $\sigma \approx 4.8 \times 10^{-2}$).
See Fig.~\ref{fig:smoothing} for an example.

The impact on model accuracy is dramatic. A network trained on the
smoothed target with only 2\,M samples achieves a median orbital-phase
endpoint error of $0.013\,\mathrm{rad}$, compared to
$0.137\,\mathrm{rad}$ for a model trained on the raw target with
5\,M samples, a $10\times$ improvement with $2.5\times$ less data.
When using the raw (un-smoothed) target, the orbital-frequency data piece accounted for
$98.2\%$ of the total time-domain mismatch. After smoothing, however, this is
reduced to $73\%$, with the remaining budget shared among the quaternion
and co-orbital modes (Sec.~\ref{sec:error_decomposition}).

Because the smoothing alters the training target itself, a
perfectly trained network would inherit a systematic difference from
\nrsur{} that a validation MSE computed against the smoothed target
cannot detect.  Two checks assess this systematic difference.  First, all
end-to-end mismatches in Sec.~\ref{sec:results} are computed against the
\emph{raw} \nrsur{} waveforms, so any conditioning-induced bias is
included in the reported distributions.  Second, we quantify the
systematic directly by assembling waveforms from ground-truth data-piece values
with the raw and the smoothed orbital phase (300 validation samples):
the TD mismatch between the two has median $3.6\times10^{-8}$ (95th
percentile $2.7\times10^{-7}$), well below the median surrogate accuracy
(Sec.~\ref{sec:e2e_accuracy}).

The smoothed target is insensitive to the window boundaries. Varying
$t_\mathrm{start}$ between $30$ and $50\,M$ leaves the target unchanged,
while varying the onset of the constant-derivative clamp, $t_\mathrm{end}$,
between $70$ and $100\,M$ changes the assembled waveform only at the
$10^{-8}$ mismatch level.

\subsection{Training procedure}
\label{sec:training}

For training, we generate 5\,000\,000 samples and a separate set of
100\,000 validation samples
by drawing parameters from the \nrsur{} domain ($q \in [1,4]$,
$|\boldsymbol{\chi}_{1,2}| \leq 0.8$) and evaluating the surrogate with
\textsc{gwsurrogate} to obtain targets for all data pieces on the sparse time
grid.  The mass ratio is drawn uniformly, and the spin components uniformly
on $[-0.8, 0.8]^3$, with any vector exceeding $|\boldsymbol{\chi}| = 0.8$
rescaled onto that sphere.  This rescaling places ${\sim}48\%$ of the spins
at maximal magnitude, so the sampled distribution is weighted toward rapidly
spinning configurations, which are the most challenging to model.
Training and validation sets are disjoint.

Using the training data, we train each network by minimizing the mean
squared error in the standardized output space,
\begin{equation}
  \mathcal{L} = \frac{1}{C\,T}
    \sum_{c=1}^{C} \sum_{k=1}^{T}
    \biggl(\frac{\hat{y}_c(t_k) - y_c(t_k)}{\sigma_c(t_k)}\biggr)^{\!2} ,
\label{eq:loss}
\end{equation}
where $\sigma_c(t_k)$ is the per-channel, per-timestep standard
deviation from the training set. Dividing by $\sigma_c(t_k)$ ensures
that all time steps contribute equally to the gradient, despite the large
dynamic range of the physical targets (Sec.~\ref{sec:smoothing}).

For the quaternion model, we add a unit-norm penalty,
\begin{equation}
  \mathcal{L}_\mathrm{quat} = \mathcal{L}_\mathrm{MSE}
    + \lambda_\mathrm{norm}
      \sum_{k=1}^{T}
      \bigl(|\hat{\mathbf{q}}(t_k)|^2 - 1\bigr)^2 \,,
\label{eq:loss_quat}
\end{equation}
with $\lambda_\mathrm{norm} = 0.01$.  This soft constraint encourages
the predicted quaternion to remain near the unit sphere; exact
normalization is enforced at assembly time (Sec.~\ref{sec:assembly}).

We optimize these losses using AdamW~\cite{Loshchilov2019} with a linear
warmup over the first 5\% of training steps,
followed by cosine annealing to zero.  The learning rate is
$10^{-3}$ for co-orbital modes and $3 \times 10^{-4}$ for dynamics
quantities at the 5\,M scale (where larger initial learning rates can cause
instabilities).  The batch size is 256 for all pieces.  Training runs
for 200 epochs at 5\,M samples and 300 epochs at 2\,M, taking 4--12
hours per piece on a single NVIDIA A100 or H100 GPU.

Finally, we assess the benefit of additional training data.  Increasing
the training set from $2\,\mathrm{M}$ to $5\,\mathrm{M}$ samples reduces
the validation MSE by $26\%$ for $\Omega$ (PieceMLP h512$\times$12),
$31\%$ for the quaternion (PieceMLP h1024$\times$6), and $73\%$ for the
$(2,2)$ co-orbital mode (PieceMLP h1024$\times$6); the modes therefore
benefit more from additional data than the dynamics quantities.
Accuracy is not saturated at $5\,\mathrm{M}$ samples, so further gains
remain available from a larger training set.  For comparison, upgrading
the quaternion network to h2048$\times$6 at $5\,\mathrm{M}$ samples
gives a further $2.2\times$ improvement over the h1024 baseline, i.e.\
capacity and data are both still binding at this scale.

\subsection{Waveform assembly}
\label{sec:assembly}

The inertial-frame strain is reconstructed from the outputs of the data-piece models in
five steps, mirroring the \nrsur{} evaluation pipeline:
\begin{enumerate}
  \item \textbf{Phase integration.}
    Integrate the predicted orbital frequency
    $\hat{\Omega}(t)$ to obtain the orbital phase.  Because the sparse
    time grid $\{t_k\}_{k=1}^{T}$ is fixed across all samples, the
    cubic-spline antiderivative is a \emph{linear operator},
    $\hat{\phi}_\mathrm{orb} = A_\phi\,\hat{\Omega}$, where the
    $T \times T$ matrix~$A_\phi$ is precomputed once from the
    not-a-knot cubic spline on the identity basis.  At runtime, phase
    integration reduces to a single matrix--vector product with no
    per-sample spline construction.
  \item \textbf{Quaternion normalization.}
    The predicted quaternion is projected onto the unit sphere according to the rule:
    $\hat{\mathbf{q}}(t) \to \hat{\mathbf{q}}(t) /
    |\hat{\mathbf{q}}(t)|$.
  \item \textbf{Full rotation.}
    The co-precessing quaternion is composed with the orbital-phase rotation using quaternion multiplication,
    $R(t) = \hat{\mathbf{q}}(t) \otimes
    R_z\bigl(\hat{\phi}_\mathrm{orb}(t)\bigr)$,
    where $R_z(\phi) = (\cos\tfrac{\phi}{2}, 0, 0,
    \sin\tfrac{\phi}{2})$ is a rotation about the $z$-axis.
  \item \textbf{Wigner $D$-rotation.}
    For each multipole $\ell$, the Wigner $D$-matrix
    $D^\ell_{m'm}(R)$ is computed using the Cayley-Klein
    parameterization~\cite{Boyle2013}.  The full rotation quaternion is
    first inverted, then decomposed into Cayley-Klein parameters
    $r_a = w + iz$ and $r_b = y + ix$.  Each matrix element is a
    polynomial in $r_a$, $\bar{r}_a$, $r_b$, and $\bar{r}_b$ with
    coefficients that depend only on $\ell$, $m'$, and $m$.  We
    precompute these coefficients once into sparse tables and evaluate
    all $(2\ell+1)^2$ elements in parallel via batched power stacks
    (all required integer powers of each Cayley-Klein parameter
    accumulated in a single vectorized pass over time steps) and
    a single matrix multiplication for the inner summation (see
    Sec.~\ref{sec:speed} for the resulting speedup).  The
    inertial-frame modes
    \begin{equation}
      h_{\ell m}^\mathrm{inert}(t) = \sum_{m'}
        D^\ell_{m'm}\bigl(R(t)\bigr)\,
        \hat{h}_{\ell m'}^\mathrm{coorb}(t) \,,
    \label{eq:wigner_rotation}
    \end{equation}
    are evaluated per~$\ell$ block, where 
    for each $\ell$ the $(2\ell+1)\times(2\ell+1)$ matrix $D^\ell$ acts
    on the vector of co-orbital modes $\hat{h}_{\ell m'}^\mathrm{coorb}$
    via a batched matrix--vector contraction over all time steps,
    rather than explicit loops over mode indices.
  \item \textbf{Strain projection.}
    The spin-weighted spherical harmonics
    ${}_{-2}Y_{\ell m}(\iota,\pi/2-\phi_\mathrm{ref})$ are precomputed
    into a 21-component vector $\mathbf{Y}$ containing the modes with
    $\ell=2,3,4$. The GW polarizations are then obtained from
    Eq.~\eqref{eq:strain} via a single contraction
    $h_+ - i h_\times = \mathbf{Y} \cdot \mathbf{h}^\mathrm{inert}$.
\end{enumerate}
All operations are differentiable and implemented as pure tensor
operations, eliminating all Python-level loops.
This enables gradient-based sensitivity analysis
(Sec.~\ref{sec:error_decomp_method}) and is compatible with
XLA compilation (JAX~\cite{jax2018github}) and \texttt{torch.compile}
(PyTorch~\cite{paszke2019pytorch,ansel2024pytorch2}).

\subsection{Parallel model bank}
\label{sec:bank}

Evaluating the models for 25 data pieces sequentially incurs significant overhead from
per-model kernel launches and memory allocation, particularly on GPUs.
To address this, we group models sharing the same 
hidden width $H$ and depth $D$ into \emph{buckets} and stack their weight matrices along
a new ``model'' dimension:
\begin{equation}
  W_\mathrm{stacked} \in \mathbb{R}^{M \times H_\mathrm{out}
  \times H_\mathrm{in}} \,,
\label{eq:stacked_weights}
\end{equation}
where $M$ is the number of models in the bucket.  A single batched
matrix multiplication (GEMM) per layer then evaluates all $M$ models
simultaneously.  Output layers, which differ in $C_i$ across models,
are handled individually but contribute negligible cost since
$H \to C_i$ is small.  Per-model de-standardization is applied using
stacked mean and standard deviation buffers.

Table~\ref{tab:bank_speedup} shows the resulting speedups.  For the
production architecture ($H = 1024$, $D = 6$, 23 models), the parallel
bank achieves $15.2\times$ speedup on GPU at batch size 1, reducing the
per-batch neural-network cost from $6.7\,\mathrm{ms}$ to
$0.44\,\mathrm{ms}$.

The speedups in Table~\ref{tab:bank_speedup} are measured for an eager
execution model whose sequential baseline launches one kernel per piece.
The JAX implementation used for the timings of Sec.~\ref{sec:speed}
instead obtains the same effect through \texttt{jax.jit}/XLA kernel fusion
and \texttt{jax.vmap} batching, without explicitly stacking the model
weights. Explicit weight stacking in JAX provides only a marginal
additional speedup (${\sim}1.1$--$1.4\times$, diminishing with batch
size), because the batched workload is limited by the efficiency of many
small per-piece GEMMs rather than by per-piece launch overhead.

\begin{table}[t]
  \centering
  \caption{Sequential evaluation vs.\ parallel model bank inference latency for models of 23 data pieces. Measured at batch size 1 on an
    A100-SXM4-80GB GPU and a CPU node.}
  \label{tab:bank_speedup}
  \begin{ruledtabular}
  \begin{tabular}{llccc}
    Device & Config & Seq.\ [ms] & Bank [ms] & Speedup \\
    \hline
    CPU  & $H{=}256$, $D{=}24$  & 20.1  & 2.9   & $6.9\times$ \\
    CPU  & $H{=}1024$, $D{=}6$  & 11.3  & 5.8   & $1.9\times$ \\
    GPU  & $H{=}256$, $D{=}24$  & 22.1  & 1.1   & $20.5\times$ \\
    GPU  & $H{=}1024$, $D{=}6$  & 6.7   & 0.44  & $15.2\times$ \\
  \end{tabular}
  \end{ruledtabular}
\end{table}

\subsection{Error decomposition}
\label{sec:error_decomp_method}

We use two complementary methods to attribute the end-to-end waveform
mismatch to errors in individual data pieces, guiding architecture selection
and identifying accuracy bottlenecks.

\subsubsection{Leave-one-out replacement}
For each evaluation waveform and data piece $i$, we assemble
$h_\mathrm{pred}$ using all NN predictions. We then construct
$h_\mathrm{patched}$ by replacing only the prediction for piece $i$
with its \nrsur{} value. The mismatch improvement is
\begin{equation}
  \Delta_i =
  \MM(h_\mathrm{pred},h_\mathrm{true})
  - \MM(h_\mathrm{patched},h_\mathrm{true}) \,,
\end{equation}
where $\Delta_i$ measures the marginal contribution of piece $i$'s error to
the total mismatch.  Since cross-terms between pieces are small when
individual errors are small, $\sum_i \Delta_i \approx
\MM_\mathrm{total}$, where $\MM_\mathrm{total}\equiv\MM(h_\mathrm{pred},h_\mathrm{true})$
is the mismatch of the fully predicted waveform against the ground truth.  The fractional contribution is
$f_i = \langle\Delta_i\rangle / \langle\MM_\mathrm{total}\rangle$,
where $\langle\cdot\rangle$ denotes averaging over evaluation samples.

\subsubsection{Monte Carlo perturbation sensitivity}
For systematic architecture optimization
(Sec.~\ref{sec:network_sizing}), we need sensitivity weights $w_i$ that
predict how a unit improvement in piece $i$'s MSE translates to
mismatch reduction, \emph{independent} of the current model's accuracy.
For each of $N$ parameter-space samples and each piece $i$, we add
independent Gaussian perturbations to its ground-truth values at every
channel and time node, reassemble the waveform, and measure the
resulting mismatch against the unperturbed truth.

Three refinements over a na\"ive fixed-$\sigma$ implementation (adding
noise with the same absolute amplitude $\sigma$ to every piece) are
required.  \emph{First}, the perturbation amplitude must be normalized
per piece.  A fixed physical $\sigma = 0.01$ represents a ${\sim}1\%$
perturbation for the dominant $(2,2)$ mode (amplitude $\mathord{\sim}1$)
but a ${\sim}10\times$ perturbation for a weak $(4,-4)$ mode (amplitude
$\mathord{\sim}10^{-3}$), creating an artificial ranking unrelated to
actual model errors.  We instead set
$\sigma_i = \sigma_0 \,\sigma^{\rm glob}_i$, where $\sigma^{\rm glob}_i$
is the global RMS amplitude of piece $i$'s output values across the
training data (the same scale factor used in output standardization
during training), and $\sigma_0 = 0.01$ is a universal dimensionless
level.  Denoting by $\MM_{\mathrm{pert},i}$ the mismatch, against the
unperturbed truth, of the waveform assembled with piece $i$'s outputs
perturbed and all other pieces held at ground truth, the weight is
reported as
\begin{equation}
  w_i = \frac{\langle\MM_{\mathrm{pert},i}\rangle}{\sigma_0^2} \,.
\label{eq:sensitivity_weight}
\end{equation}
With this normalization, $w_i$ is the mismatch incurred per unit of
normalized variance. Given the standardized validation MSE
$\varepsilon_i^2$ of piece $i$ (as reported by the training pipeline),
the product $w_i\,\varepsilon_i^2$ estimates the mismatch contribution
of that piece to the full waveform error.

\emph{Second}, because the projection from inertial-frame modes to
strain depends on the viewing inclination, we average the sensitivity
weights over $n_\iota=11$ inclinations uniformly spaced in $[0,\pi]$.
This avoids basing the mode sensitivities on a single viewing angle.  Using an odd $n_\iota$ ensures that
$\iota = \pi/2$ is included exactly, where $m=0$ modes attain their
maximum projection weight.

\emph{Third}, the orbital phase model is trained to predict
$\Omega(t) = \dd\phi_\mathrm{orb}/\dd t$, not $\phi_\mathrm{orb}$
directly (Sec.~\ref{sec:tcmlp}). Perturbing $\phi_\mathrm{orb}$ directly would produce a sensitivity
weight for a quantity that the network does not predict and for which
no accuracy profile exists. Instead, we perturb $\Omega$ at each node with noise
$\delta\Omega_j \sim \mathcal{N}(0,\sigma_\Omega^2)$ and integrate
forward to obtain the phase perturbation
\begin{equation}
  \delta\phi(t_k) = \sum_{j=0}^{k-1} \delta\Omega_j\,(t_{j+1}-t_j),
\end{equation}
which is added to the unperturbed $\phi_\mathrm{orb}$ before reassembly.
Note that this forward-Euler sum integrates only the stochastic
perturbation $\delta\Omega$. The baseline $\phi_\mathrm{orb}$ is the
exact ground truth, and the assembled waveform uses the same
cubic-spline antiderivative as production (Sec.~\ref{sec:assembly}).
Because $\delta\Omega$ is white noise, the accumulated $\delta\phi$ is
largely insensitive to the choice of integrator. For 
example, the forward Euler and the
cubic-spline operator agree to better than $3\%$ per node and are
identical in root-mean-square, so the simpler scheme suffices here.
The resulting weight measures the waveform sensitivity to errors in
the modeled $\Omega$, including their accumulation during the time
integration used to assemble the waveform.

We use $N = 200$ samples, $K = 5$ perturbations per piece per sample,
and $n_\iota = 11$ inclinations.  Since the noise is injected into ground-truth outputs for the data pieces rather than
network predictions, the networks are never invoked during this
procedure.  The weights $w_i$ therefore depends only on how the assembly pipeline
propagates errors into mismatch and on the parameter-space distribution
of the samples.  It is an intrinsic property of the waveform family,
computed once and reused across all architecture comparisons.

\subsection{Error-budget-driven network sizing}
\label{sec:network_sizing}

For each data piece $i$, let $\mathcal{C}_i$ denote the 
set of trained candidate configurations and let
$c_i\in\mathcal{C}_i$ be the selected configuration. A configuration
specifies the architecture family and its hyperparameters, including
the hidden width $H$ and depth $D$. Each candidate has a measured
standardized validation MSE $\varepsilon_i^2(c_i)$ and a trainable
parameter count $P(c_i)$. We choose one configuration per data piece
to minimize the total predicted mismatch under an illustrative total
parameter-count budget $B=50\,\mathrm{M}$:
\begin{equation}
  \begin{aligned}
    \min_{\{c_i\in\mathcal{C}_i\}}\quad&
      \sum_{i=1}^{N_\mathrm{pieces}} w_i\,\varepsilon_i^2(c_i),\\
    \text{subject to}\quad&
      \sum_{i=1}^{N_\mathrm{pieces}} P(c_i) \leq B .
  \end{aligned}
\label{eq:knapsack}
\end{equation}
Here $w_i$ are the sensitivity weights
(Sec.~\ref{sec:error_decomp_method}) and $\varepsilon_i^2(c_i)$ is the
validation MSE of piece $i$ at configuration $c_i$, computed in the
standardized output space (i.e., before de-standardization by
$\sigma_c(t_k)$, as in the training loss Eq.~\eqref{eq:destandardize}).
Because $w_i$ is measured in the same normalized units as
$\varepsilon_i^2$ (by construction of the per-piece $\sigma_i$, see
Eq.~\eqref{eq:sensitivity_weight}), the product $w_i \varepsilon_i^2$
is a consistent predictor of the mismatch contribution regardless of
each piece's physical amplitude scale.
Eq.~\eqref{eq:knapsack} is a multiple-choice knapsack problem~\cite{Kellerer2004}, with one configuration selected from each set $\mathcal{C}_i$.

The optimization combines sensitivity, accuracy, and cost information.
For each data piece, the sensitivity weight $w_i$ is computed as
described in Sec.~\ref{sec:error_decomp_method}. The accuracy profile
$\varepsilon_i^2(c)$ is measured by training the candidates
$c\in\mathcal{C}_i$. Candidate sets may differ among data pieces. They
include PieceMLPs with different widths and depths, as well as
specialized models such as TCMLP+SC for $\Omega$.

The remaining choice is how to measure candidate cost and set the
budget. Inference latency is not additive for the parallel model bank
because models with the same architecture are evaluated together. For
$128\leq H\leq512$, adding models changes the total GPU latency very
little. A bank of 25 such models takes approximately $0.22$\,ms,
whereas a 25-model bank with $H=1024$ takes $0.72$\,ms. Summing
standalone model latencies therefore overestimates the measured bank
latency by approximately $30\times$ for the smaller architectures. We
summarize this scaling using the average latency added per model,
\begin{equation}
  \overline{\Delta t}_\mathrm{bank}
  =
  \frac{t_\mathrm{bank}^{(N)}-t_\mathrm{bank}^{(1)}}{N-1},
\end{equation}
where $t_\mathrm{bank}^{(N)}$ is the latency for a bank of $N$ models.
This increment is approximately $0.01$\,ms for $H\leq512$ and
$0.02$\,ms for $H=1024$.

The complete bank takes at most $0.72$\,ms for the tested candidates,
so GPU latency does not meaningfully constrain the selection. We
therefore use the total parameter count $P(c)$ as the cost in
Eq.~\eqref{eq:knapsack}. Parameter count is device independent and
provides a simple proxy for training cost. The illustrative budget
$B=50\,\mathrm{M}$ is larger than the $36\,\mathrm{M}$ parameters
required by the all-$H=512$ bank, but smaller than the
$137\,\mathrm{M}$ required by the all-$H=1024$ bank. It therefore
allows some data pieces to use wider networks without assigning them
to the full bank.

\paragraph{Greedy optimizer.}
We solve Eq.~\eqref{eq:knapsack} greedily: initialize each data
piece with the candidate in $\mathcal{C}_i$ having the smallest
parameter count, then iteratively upgrade the piece with the highest
benefit per additional parameter,
\begin{equation}
  r_i = \frac{w_i \,\bigl[\varepsilon_i^2(c_i^\mathrm{curr})
              - \varepsilon_i^2(c_i^\mathrm{next})\bigr]}
             {P(c_i^\mathrm{next}) - P(c_i^\mathrm{curr})}
\end{equation}
until the budget is exhausted or no positive-benefit upgrade
remains. Here $c_i^\mathrm{next}$ is the next configuration in order of
decreasing MSE, so the numerator is non-negative by construction. When
that configuration is also no more expensive (non-positive
denominator), the upgrade is dominating and is taken first.
We use this greedy rule as a practical heuristic for the discrete
optimization problem in Eq.~\eqref{eq:knapsack}.

The allocation results with the full corrected methodology are reported
in Sec.~\ref{sec:network_sizing_results}.

\section{Results}
\label{sec:results}

\subsection{Accuracy of individual data pieces}
\label{sec:piece_accuracy}

\begin{figure}[t]
  \centering
  \includegraphics[width=\columnwidth]{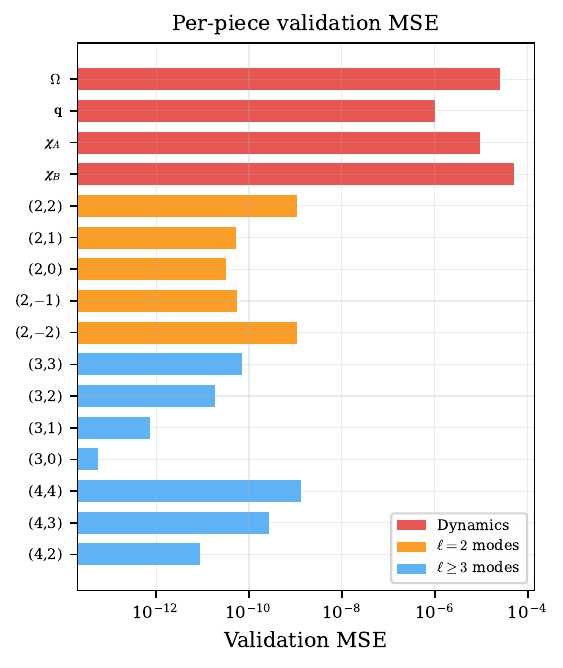}
  \caption{Standardized validation MSE $\varepsilon_i^2(c_i)$
    (Sec.~\ref{sec:network_sizing}) for each data piece at its selected
    architecture configuration, from the error-budget-driven network sizing run, which
    closely matches the production bank (Table~\ref{tab:piece_summary}).
    The dynamics quantities (quaternion, $\Omega$, spins) dominate
    the error budget, whereas co-orbital modes are fitted 3--6 orders of
    magnitude more accurately.}
  \label{fig:piece_mse}
\end{figure}

Table~\ref{tab:piece_summary} and Fig.~\ref{fig:piece_mse} summarize the architecture and validation
MSE for each data piece in the production model bank.  The error hierarchy
spans more than six orders of magnitude. Co-orbital modes achieve
$\mathrm{MSE} \sim 10^{-9}$, while dynamics quantities (quaternion,
orbital frequency, spin trajectories) range from $10^{-6}$ to
$10^{-5}$.

We find that the co-orbital modes are well-captured by relatively small networks
($H = 256$, $D = 4$ for $\ell \geq 3$; $H = 1024$, $D = 6$ for the
dominant $\ell = 2$ modes).  This is expected because the modes, when
expressed in the co-orbital frame, are smooth, slowly varying functions
of the binary parameters, with most of the complex dynamics absorbed
into the quaternion and orbital phase.

Among the dynamics quantities, the quaternion is the most challenging
because it encodes the full precession dynamics and enters the Wigner-$D$
rotation that affects all 21 modes.  A wide network ($H = 2048$, $D = 6$)
with 23\,M parameters is required to achieve $\mathrm{MSE} = 1.02 \times
10^{-6}$.  The orbital frequency, trained on the smoothed target
(Sec.~\ref{sec:smoothing}), reaches $\mathrm{MSE} = 2.55 \times 10^{-5}$
with the TCMLP+SC (Sec.~\ref{sec:tcmlp}).

The spin trajectories show an asymmetry. Using the same PieceMLP architecture and training setup, the Cartesian-component validation MSE for $\boldsymbol{\chi}_B$ is $5.4$ larger than $\boldsymbol{\chi}_A$. One potential explanation is that the lighter black hole's spin is more sensitive to the coupling with the orbital dynamics at high mass ratios. However, a counter effect is that at unequal mass-ratios the secondary spin has less impact on the waveform, so these effects may offset.

\begin{table}[t]
  \centering
  \caption{Architecture and validation MSE for each data piece in the
    production model bank. All models are trained on the sparse \nrsur{}
    grid ($T \approx 230$). Here $H$ is the hidden-layer width, $D$ is
    the number of hidden layers, $C$ is the number of output channels,
    and Val MSE is the dimensionless mean squared error of the
    standardized network outputs on the validation set. For each mode
    group, $\leq$ marks an upper bound set by the largest MSE among its
    constituent modes.}
  \label{tab:piece_summary}
  \begin{ruledtabular}
  \begin{tabular}{lllrrl}
    Data piece & Arch & $H \times D$ & $C$ & Params & Val MSE \\
    \hline
    $\Omega$              & TCMLP+SC & $384 \times 16$ & 1 & 2.24M & $2.55 \times 10^{-5}$ \\
    $\mathbf{q}$          & PieceMLP & $2048 \times 6$ & 4 & 22.9M & $1.04 \times 10^{-6}$ \\
    $\boldsymbol{\chi}_A$ & PieceMLP & $1024 \times 6$ & 3 &  6.3M & $9.20 \times 10^{-6}$ \\
    $\boldsymbol{\chi}_B$ & PieceMLP & $1024 \times 6$ & 3 &  6.3M & $4.96 \times 10^{-5}$ \\
    \hline
    $\ell{=}2$ modes & PieceMLP & $512 \times 6$ & 2 & 1.55M & ${\leq}\,3.1\times10^{-9}$ \\
    $\ell{=}3$ modes & PieceMLP & $512 \times 6$ & 2 & 1.55M & ${\leq}\,2\times10^{-10}$  \\
    $\ell{=}4$ modes & PieceMLP & $256 \times 4$ & 2 &  0.3M & ${\leq}\,6\times10^{-9}$   \\
  \end{tabular}
  \end{ruledtabular}
\end{table}

\subsection{End-to-end waveform accuracy}
\label{sec:e2e_accuracy}

We evaluate the complete neural surrogate against \nrsur{} on 10\,000
validation waveforms drawn from the same parameter distribution as the
training set (Sec.~\ref{sec:training}).  The
waveform is assembled from predictions for the data pieces as described in
Sec.~\ref{sec:assembly} and compared to the ground truth using the
mismatch metrics of Sec.~\ref{sec:mismatch_defs}.

\begin{figure}[t]
  \centering
  \includegraphics[width=\columnwidth]{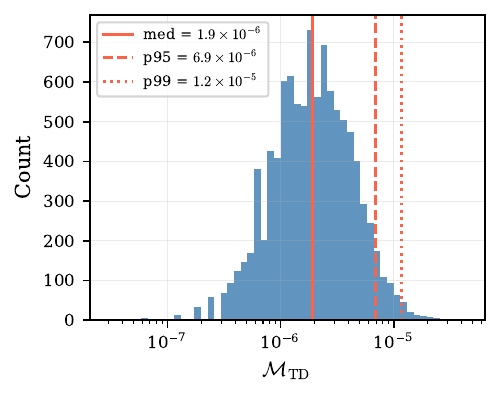}
  \caption{Time-domain mismatch distribution ($M = 120\,\Msun$,
    face-on) over 10\,000 evaluation waveforms, with percentile markers.
    Although this evaluation was run with $M=120\,\Msun$, the
    dimensionless TD mismatch is independent of total mass, so the same
    distribution applies at any total mass.}
  \label{fig:td_mismatch}
\end{figure}

We first discuss the time-domain mismatch, a useful diagnostic that
we also use for the error decomposition and piece-swap analyses below.
To isolate the errors of $\Omega$, the quaternion, and the co-orbital
modes from those of the spin networks, this diagnostic feeds the
\nrsur{} ground-truth spin trajectories, rather than the predicted ones,
into the spin-conditioned $\Omega$ network.  Because $\Omega$ is the
only spin-conditioned network and the spins enter the waveform only
through it, this removes the spin-network error from the diagnostic
entirely; the fully end-to-end configuration with predicted spins is
used for the frequency-domain results below.
At face-on inclination ($\iota = 0$), the TD mismatch over 10\,000
validation waveforms has a median of $1.91 \times 10^{-6}$, a 95th
percentile of $6.85 \times 10^{-6}$, and a 99th percentile of
$1.17 \times 10^{-5}$; see Fig.~\ref{fig:td_mismatch} for the full
distribution.

\begin{figure}[t]
  \centering
  \includegraphics[width=\columnwidth]{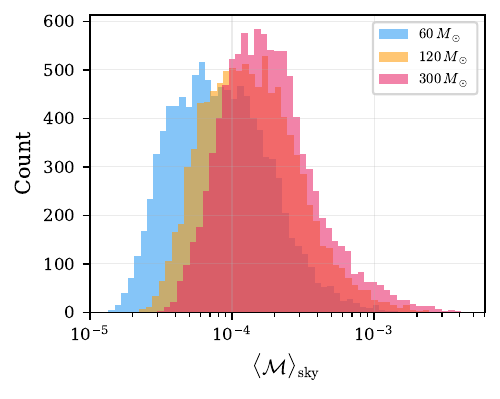}
  \caption{Sky-averaged FD mismatch distributions at three total masses.
    The mismatch shifts to higher values with increasing $M$ as the
    merger-ringdown band becomes more prominent.  Even at
    $M = 300\,\Msun$, the median remains below $2 \times 10^{-4}$.}
  \label{fig:fd_mismatch}
\end{figure}

\begin{figure}[t]
  \centering
  \includegraphics[width=\columnwidth]{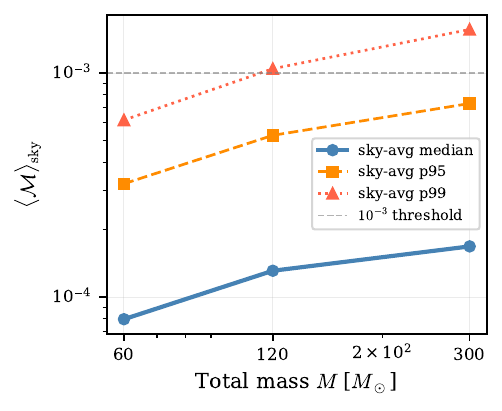}
  \caption{Sky-averaged FD mismatch statistics as a function of total mass $M$.
The dashed line shows the commonly used criterion
$\MM=1/(2\rho^2)$, where $\rho$ is the signal-to-noise ratio
(Sec.~\ref{sec:e2e_accuracy}). For $\rho=22$, this corresponds to
$\MM\approx10^{-3}$. Waveform differences below this value are
generally considered unresolvable at that signal-to-noise ratio. The
median and 95th percentile remain below the threshold across the full
mass range.}
  \label{fig:fd_vs_mass}
\end{figure}

Next we turn to the sky-averaged frequency-domain mismatch, which is the
primary metric for GW data analysis.  These results are produced by the
full JAX production model evaluated end to end.
Table~\ref{tab:fd_mismatch} and Fig.~\ref{fig:fd_mismatch} report the
sky-averaged and worst-sky FD mismatch at three total masses. The
mismatch increases with total mass, consistent with the larger fraction
of the in-band signal contributed by merger and ringdown (see also
Fig.~\ref{fig:fd_vs_mass}). At the largest mass considered,
$M=300\,\Msun$, the 95th percentile of the sky-averaged mismatch is
$7.32 \times 10^{-4}$.
For context, a commonly used reference criterion for waveform
indistinguishability is $\MM < 1/(2\rho^2)$, where $\rho$ is the
signal-to-noise ratio~\cite{Lindblom:2008cm}.  For a typical loud event with $\rho = 20$,
this gives $\MM < 1.25 \times 10^{-3}$; our sky-averaged median
at $M = 120\,\Msun$ ($1.31 \times 10^{-4}$) satisfies this with a
factor of $\sim 10$ margin, and the sky-averaged 95th percentile
($5.27 \times 10^{-4}$) still lies comfortably below the threshold.
Moreover, the \nrsur{} model itself has
a median TD mismatch of $\sim 10^{-4}$ against the underlying NR
simulations~\cite{Varma:2019csw}; the neural network adds only
$\sim 2 \times 10^{-6}$ (TD median), nearly two orders of magnitude
smaller than the intrinsic model error.

All mismatches reported in this section were computed with
the data-piece networks in true IEEE single precision and the orbital-phase
integration, Wigner-$D$ rotation, and mismatch integrals in double
precision -- so they are therefore free of TF32 effects.  We checked directly
on an L40S GPU that this does not matter.  The true-single-precision GPU
evaluation reproduces the CPU values to ${\lesssim}3\times10^{-8}$, at
the level of the single-precision reduction-order differences between the
two backends (the networks run in float32, so the same GEMMs accumulate
in a different order on GPU and CPU), and forcing the batched network
evaluation to TF32 instead shifts the sky-averaged mismatch by only
${\sim}10^{-7}$ in the median (at most a few${\times}10^{-6}$ across
500~waveforms), leaving the distribution statistics of
Table~\ref{tab:fd_mismatch} unchanged at the quoted precision.  This is
the same TF32 arithmetic used for eligible matrix multiplications in the
batched GPU timings of Sec.~\ref{sec:speed}. The corresponding precision
and performance tradeoffs are discussed in
Sec.~\ref{sec:implementations}.

\begin{table}[t]
  \centering
  \caption{Frequency-domain mismatches computed using
the aLIGOLateHighSensitivity PSD and waveform-dependent frequency bounds.
Results are based on 10\,000 evaluation waveforms evaluated at 27
directions on the sphere. \emph{Sky avg.} denotes the average over these
directions, while \emph{Worst sky} denotes the maximum.}
  \label{tab:fd_mismatch}
  \begin{ruledtabular}
  \begin{tabular}{llccc}
    $M\;[\Msun]$ & Sky & Median & 95th pctile & 99th pctile \\
    \hline
    \multirow{2}{*}{$60$}
      & Sky-avg   & $7.96 \times 10^{-5}$ & $3.21 \times 10^{-4}$ & $6.17 \times 10^{-4}$ \\
      & Worst sky & $1.28 \times 10^{-4}$ & $5.10 \times 10^{-4}$ & $9.95 \times 10^{-4}$ \\
    \multirow{2}{*}{$120$}
      & Sky-avg   & $1.31 \times 10^{-4}$ & $5.27 \times 10^{-4}$ & $1.05 \times 10^{-3}$ \\
      & Worst sky & $2.30 \times 10^{-4}$ & $9.01 \times 10^{-4}$ & $1.76 \times 10^{-3}$ \\
    \multirow{2}{*}{$300$}
      & Sky-avg   & $1.68 \times 10^{-4}$ & $7.32 \times 10^{-4}$ & $1.57 \times 10^{-3}$ \\
      & Worst sky & $3.16 \times 10^{-4}$ & $1.38 \times 10^{-3}$ & $2.96 \times 10^{-3}$ \\
  \end{tabular}
  \end{ruledtabular}
\end{table}

\subsection{Error decomposition}
\label{sec:error_decomposition}

\begin{figure*}[t]
  \centering
  \includegraphics[width=\textwidth]{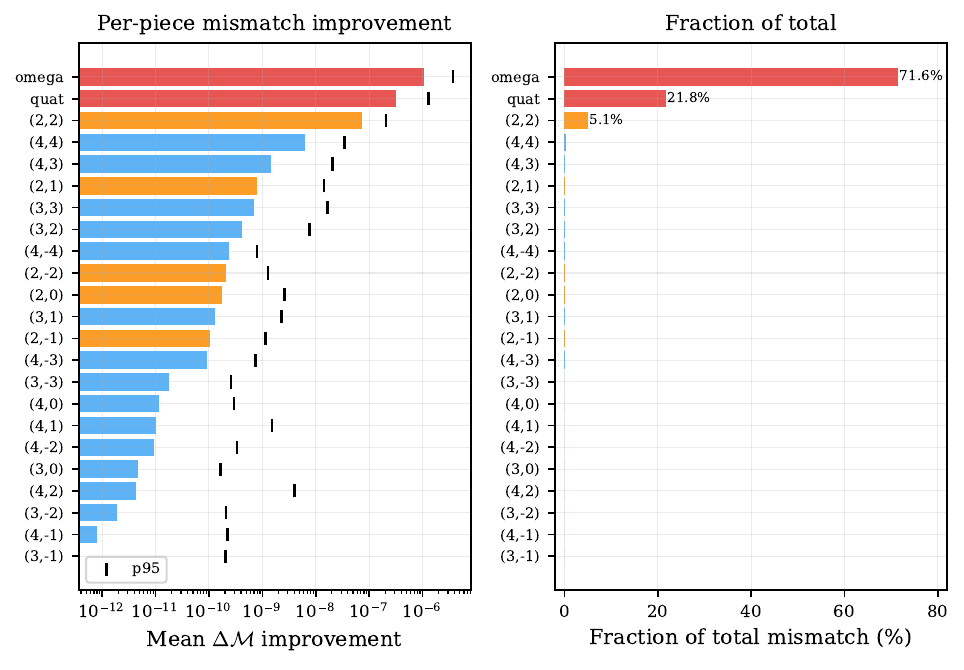}
  \caption{Leave-one-out decomposition of the TD mismatch by data piece. The left
panel shows the mean decrease in mismatch obtained by replacing each
prediction in turn with its corresponding \nrsur{} value. Markers show
the 95th percentile. The right panel shows the fraction of the total
mismatch attributed to each data piece. Bar colors follow
Fig.~\ref{fig:piece_mse}.
  }
  \label{fig:error_decomp}
\end{figure*}

Applying the leave-one-out replacement method
(Sec.~\ref{sec:error_decomp_method}) to the 10\,000 evaluation
waveforms reveals which data pieces limit the surrogate's accuracy.
Figure~\ref{fig:error_decomp} shows the mean mismatch improvement
$\langle\Delta_i\rangle$ when the prediction for each data piece is
replaced with ground truth,
and the corresponding fraction of total mismatch.

The orbital frequency $\Omega$ is the single largest error source,
accounting for $71.6\%$ of the total TD mismatch.  Phase errors in $\Omega$
accumulate over the 230-node time grid into growing orbital-phase
offsets that shift every co-orbital mode simultaneously, making this
the dominant bottleneck.  The quaternion is the second largest
contributor at $21.8\%$, reflecting its role in the Wigner-$D$ rotation
that affects all 21 inertial-frame modes.  Together, $\Omega$ and
$\mathbf{q}$ account for $93.4\%$ of the total mismatch.  The
$(2,2)$ co-orbital mode contributes $5.1\%$; all other modes are negligible.

The relative importance of $\Omega$ and $\mathbf{q}$ changes across
parameter space. Although $\Omega$ is dominant overall, the quaternion
takes over primarily at the highest mass ratios, with a weaker tendency
toward large $\chip$, where precession is strong and quaternion errors
are largest.

\begin{figure}[t]
  \centering
  \includegraphics[width=\columnwidth]{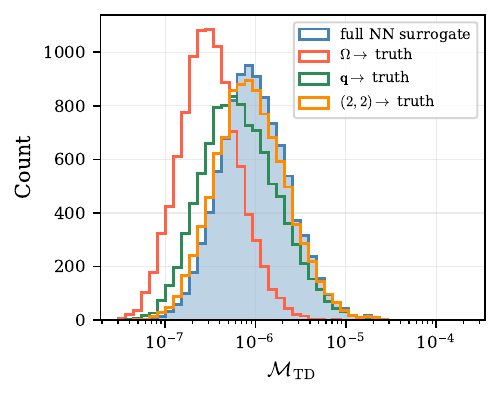}
  \caption{Effect of errors in individual data pieces on the end-to-end
    time-domain mismatch, over the 10\,000 evaluation waveforms at
    face-on inclination.  Each coloured curve replaces a single
    data-piece prediction with its \nrsur{} ground truth before assembly,
    keeping all others at their neural-network values; the horizontal
    separation from the full-surrogate distribution (blue) measures how
    much that data piece contributes to the error budget.  The orbital
    frequency $\Omega$ carries by far the largest share, the quaternion
    a smaller one, and the dominant $(2,2)$ co-orbital mode almost none,
    in agreement with Fig.~\ref{fig:error_decomp}.}
  \label{fig:piece_swap}
\end{figure}

The same hierarchy appears in the mismatch distributions shown in
Fig.~\ref{fig:piece_swap}, following earlier NR-surrogate validations.
Replacing the predicted $\Omega$ with its ground-truth value reduces
the median TD mismatch from $9.4\times10^{-7}$ to
$3.0\times10^{-7}$. The 95th percentile similarly falls from
$4.4\times10^{-6}$ to $1.1\times10^{-6}$. Replacing the quaternion or
the dominant $(2,2)$ co-orbital mode produces smaller changes, giving
median mismatches of $6.4\times10^{-7}$ and $8.7\times10^{-7}$,
respectively. These distributions are consistent with the fractional
attribution above.

\begin{figure}[t]
  \centering
  \includegraphics[width=\columnwidth]{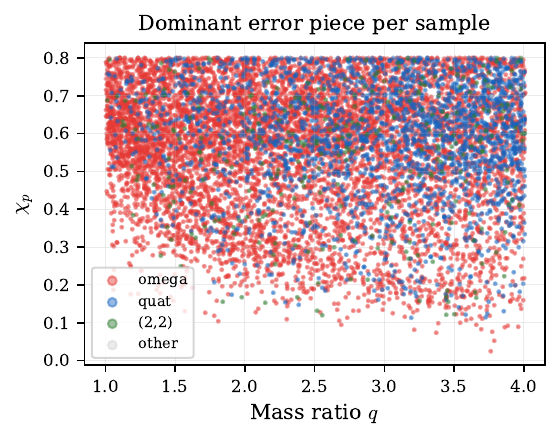}
  \caption{Data piece with the dominant error contribution per waveform
    in the $(q, \chip)$ plane.  The orbital frequency $\Omega$ (red)
    dominates across most of parameter space; the quaternion (blue)
    takes over primarily at the highest mass ratios, with a weaker
    tendency toward large $\chip$.}
  \label{fig:dominant_piece}
\end{figure}

\subsection{Error-budget-driven network sizing}
\label{sec:network_sizing_results}

Table~\ref{tab:sensitivity_weights} reports the sensitivity weights
$w_i$ obtained from the Monte Carlo perturbation procedure
(Sec.~\ref{sec:error_decomp_method}) using 200 parameter-space
samples, 5 perturbations per piece per sample, and 11 inclination
angles, together with the greedy sizing allocation at
$B = 50\,\mathrm{M}$~parameters.

\begin{table*}[t]
  \centering
  \caption{Sensitivity weights $w_i$, their fractional contributions
    $w_i/w_\mathrm{tot}$, and configurations selected by the
    error-budget-driven sizing procedure of
    Sec.~\ref{sec:network_sizing} under the parameter-count budget
    $B=50\,\mathrm{M}$. The final column reports the validation MSE used
    in the sizing calculation. The dagger marks the TCMLP+SC model used
    for $\Omega$ (Sec.~\ref{sec:tcmlp}). All other models use PieceMLP.
    The selected bank has
    $w_\mathrm{tot}=4.99$, uses $47.6\,\mathrm{M}$ parameters
    ($95\%$ of $B$), and attains an objective value
    $\sum_i w_i\,\varepsilon_i^2(c_i) = 1.2\times10^{-4}$, the predicted
    mismatch minimized in Eq.~\eqref{eq:knapsack}.}
  \label{tab:sensitivity_weights}
  \begin{ruledtabular}
  \begin{tabular}{lrrlr}
    Data piece & $w_i$ & Frac. & $H\times D$ & MSE \\
    \hline
    $\Omega$                    & $4.80$              & 96.2\%      & $384{\times}16^\dagger$ & $2.6\times10^{-5}$ \\
    $\mathbf{q}$                & $0.178$             &  3.6\%      & $2048{\times}6$         & $1.0\times10^{-6}$ \\
    $(2,-2)$                    & $4.97\times10^{-3}$ &  0.10\%     & $512{\times}6$          & $2.8\times10^{-9}$ \\
    $(2,+2)$                    & $4.89\times10^{-3}$ &  0.10\%     & $512{\times}6$          & $3.1\times10^{-9}$ \\
    $(3,-3)$                    & $2.46\times10^{-4}$ & ${<}0.01\%$ & $512{\times}6$          & $2.0\times10^{-10}$ \\
    $(3,+3)$                    & $2.40\times10^{-4}$ & ${<}0.01\%$ & $512{\times}6$          & $1.8\times10^{-10}$ \\
    $(2,\pm1)$                  & $1.4\times10^{-4}$  & ${<}0.01\%$ & $512{\times}6$          & $1.4\times10^{-10}$ \\
    Other modes                 & $\approx0$          & ---         & $256{\times}4$--$512{\times}6$ & --- \\
    $\boldsymbol{\chi}_{A,B}$   & $0$                 & ---         & $512{\times}6$          & --- \\
  \end{tabular}
  \end{ruledtabular}
\end{table*}

The sensitivity budget is dominated by $\Omega$ and $\mathbf{q}$,
which together account for ${\approx}99.8\%$ of the total. This reflects
how their errors propagate through waveform assembly. Errors in
$\Omega$ accumulate in the orbital phase, while quaternion errors
affect the rotation of every mode. Co-orbital-mode errors remain local
in time and contribute less than $0.3\%$ in total. The spin trajectories
have zero direct weight because they enter the waveform only through
the spin-conditioned $\Omega$ model.

The sensitivity weights confirm that $\Omega$ and $\mathbf{q}$ together
account for ${\approx}99.8\%$ of the total, with all co-orbital modes
combined below $0.3\%$.  The greedy allocator therefore concentrates the
budget on the two dynamics quantities.

The quaternion alone consumes 22.9\,M~parameters (48\% of $B$),
reflecting its high sensitivity weight and the accuracy gain from
scaling to $H{=}2048$: the $2048{\times}6$ model trained at 5\,M
samples achieves $\mathrm{MSE}{=}1.02{\times}10^{-6}$, a $2.2{\times}$
improvement over the $1024{\times}6$ baseline.
The $\Omega$ data piece is assigned the best available TCMLP+SC configuration
($384{\times}16$, 2.3\,M~parameters; Sec.~\ref{sec:tcmlp}), which achieves $\mathrm{MSE}{=}2.55{\times}10^{-5}$
on the smoothed target.  Zero-weight modes are assigned the cheapest
available configuration ($256{\times}4$ or $512{\times}6$) without
impacting the weighted error.

\subsection{Waveform comparisons}
\label{sec:waveform_examples}

To illustrate the quality of the neural surrogate in concrete terms, we
compare it to \nrsur{} on a representative example from the challenging
high-mass-ratio, high-spin region of the parameter space. The example
has $q = 3.5$,
$\boldsymbol{\chi}_1 = (0.7, 0.0, 0.3)$,
$\boldsymbol{\chi}_2 = (-0.1, 0.0, 0.2)$,
inclination $\iota = \pi/3$, and reference phase $\phi_\mathrm{ref} = 0.2\,\mathrm{rad}$.
At these parameters, the flat time-domain mismatch is
$\MM_\mathrm{TD} = 6.4 \times 10^{-5}$ and the sky-averaged
FD mismatch at $M = 60\,\Msun$ is $3.8 \times 10^{-4}$.

\begin{figure*}[t]
  \centering
  \includegraphics[width=\textwidth]{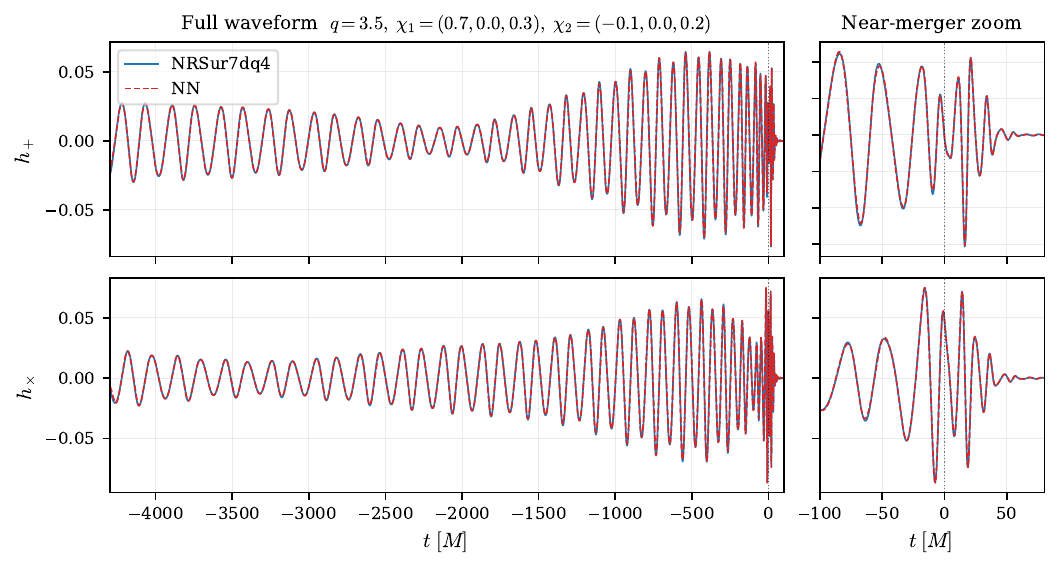}
  \caption{Comparison of the strain polarizations for a binary with
    $q = 3.5$,
    $\boldsymbol{\chi}_1 = (0.7, 0.0, 0.3)$,
    $\boldsymbol{\chi}_2 = (-0.1, 0.0, 0.2)$,
    $\iota = \pi/3$, and
    $\phi_\mathrm{ref} = 0.2\,\mathrm{rad}$. The upper and lower rows
    show $h_+(t)$ and $h_\times(t)$, respectively. Solid blue curves
    show \nrsur{}, and dashed red curves show the neural surrogate.
    The left column spans the full waveform, including the
    ${\sim}4300\,M$ inspiral. The right column spans the interval from
    ${\sim}100\,M$ before to ${\sim}80\,M$ after merger, defined as
    $t=0$ (vertical dotted line). The flat TD mismatch for this example is
    $\MM_\mathrm{TD} = 6.4 \times 10^{-5}$.}
  \label{fig:nn_waveform}
\end{figure*}

Figure~\ref{fig:nn_waveform} compares $h_+(t)$ and $h_\times(t)$ over
the full waveform, including the ${\sim}4300\,M$ inspiral, and in a
near-merger zoom. The two waveforms are visually indistinguishable on
the scale of the full waveform. Differences become visible only in the
zoom window, where a small phase offset of order
$10^{-2}\,\mathrm{rad}$ accumulates near merger.

\begin{figure*}[t]
  \centering
  \includegraphics[width=.8\textwidth]{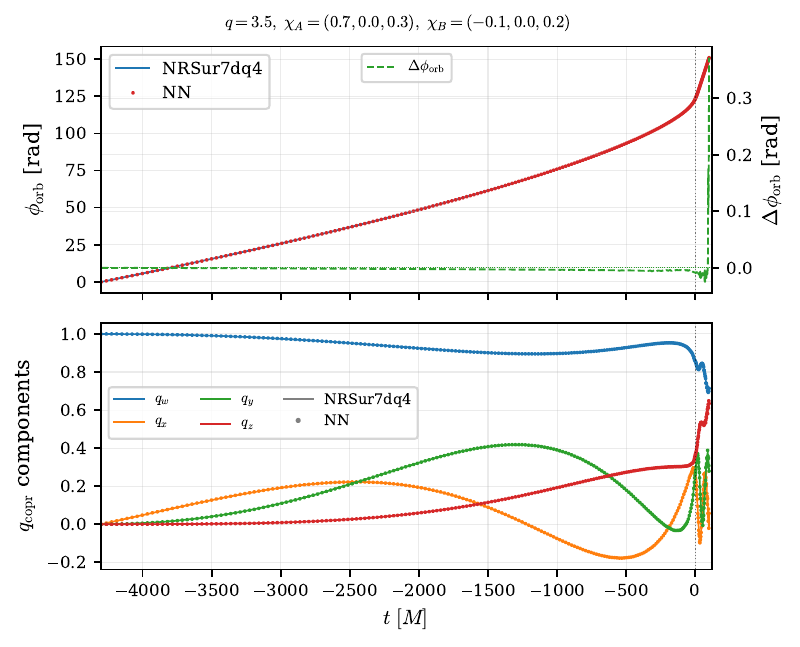}
  \caption{Orbital phase and co-precessing quaternion for the binary in
    Fig.~\ref{fig:nn_waveform}. Solid curves show \nrsur{}, and points
    show the neural surrogate on its native grid of 230 time samples. The
    upper panel shows $\phi_\mathrm{orb}(t)$ on the left axis. The residual
    $\Delta\phi_\mathrm{orb}
    =\phi_\mathrm{orb}^\mathrm{NN}-\phi_\mathrm{orb}^{\nrsur{}}$ is
    shown as a dashed green curve on the right axis. The lower panel
    shows the four components of $\mathbf{q}_\mathrm{copr}(t)$. The
    vertical dotted line marks merger, $t=0$.}
  \label{fig:nn_dynamics}
\end{figure*}

Figure~\ref{fig:nn_dynamics} compares the internal dynamics produced
by the neural surrogate with those of \nrsur{}. The top panel shows the
orbital phase $\phi_\mathrm{orb}(t)$, which grows to several hundred
radians over the inspiral. The residual
$\Delta\phi_\mathrm{orb} = \phi_\mathrm{orb}^\mathrm{NN} - \phi_\mathrm{orb}^\mathrm{\nrsur{}}$,
shown on the right axis, remains below $0.1\,\mathrm{rad}$ throughout
the coalescence and rises only deep in the ringdown, where the waveform
amplitude has decayed significantly. The bottom panel shows the four
components of the co-precessing quaternion
$\mathbf{q}_\mathrm{copr}(t)$. The neural surrogate values on the native
230-point grid follow the \nrsur{} curves closely, with the largest
deviations appearing near merger, where the precession dynamics are
most rapid. The growth of $\Delta\phi_\mathrm{orb}$ in the ringdown
partly reflects the deliberate smoothing and late-time clamping of the
$\Omega$ target described in Sec.~\ref{sec:smoothing}, rather than
neural approximation error alone. No analogous preprocessing is applied
to the quaternion, so its residual remains part of the surrogate error,
although its effect on the waveform decreases with the ringdown amplitude.

\begin{figure*}[t]
  \centering
  \includegraphics[width=\textwidth]{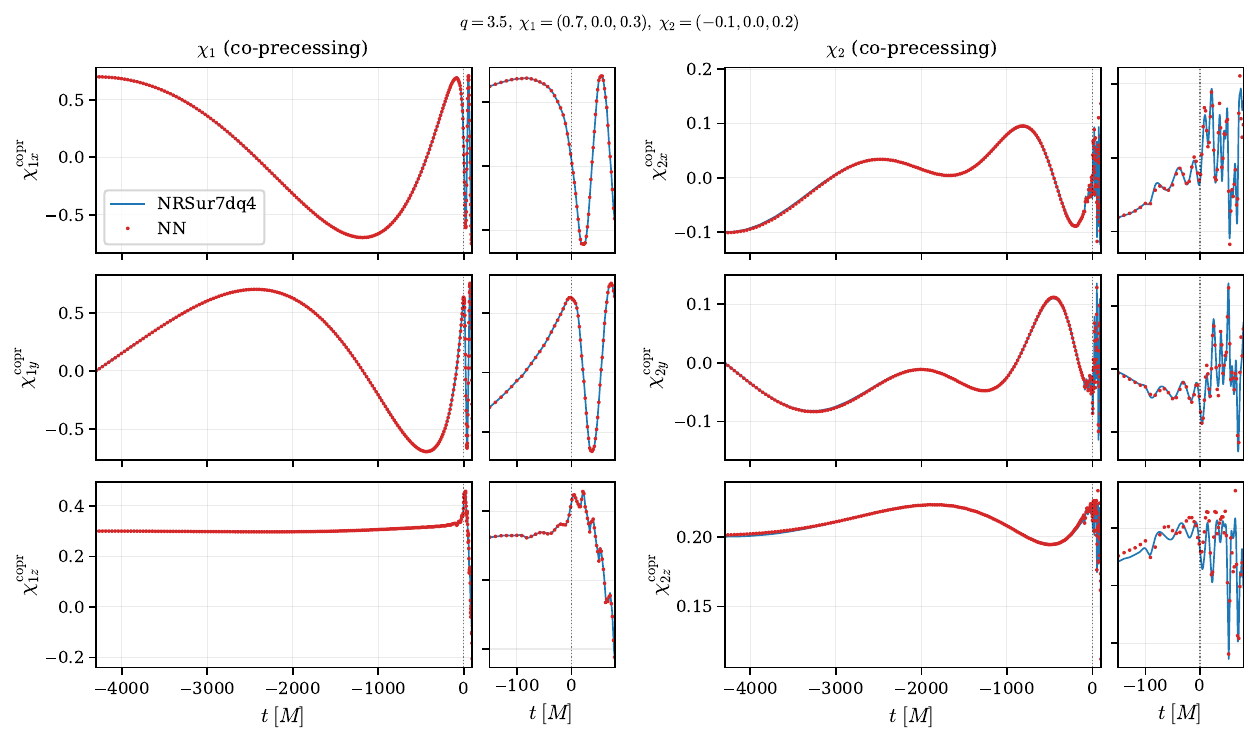}
  \caption{Co-precessing-frame spin trajectories for the binary in
    Fig.~\ref{fig:nn_waveform}.
    The upper, middle, and lower rows show the $x$, $y$, and $z$
    components, respectively. The left pair of columns shows
    $\boldsymbol{\chi}_1^\mathrm{copr}$, and the right pair shows
    $\boldsymbol{\chi}_2^\mathrm{copr}$. Within each pair, the wider
    panel spans the full waveform and the narrower panel spans the
    interval from ${\sim}150\,M$ before to ${\sim}80\,M$ after the
    merger, $t=0$ (vertical dotted line). Solid blue curves show \nrsur{}, and
    red points show the neural surrogate on its native grid of 230 time
    samples.}
  \label{fig:nn_spins}
\end{figure*}

Figure~\ref{fig:nn_spins} compares the spin trajectories
$\boldsymbol{\chi}_{1,2}^\mathrm{copr}(t)$ in the co-precessing
frame. All three Cartesian components of both spins agree well
throughout the inspiral. The largest differences in the secondary-spin
components occur near merger. For $t > -100\,M$, however, the
apparent-horizon spin measurements become unreliable as the horizons
grow highly distorted, and \nrsur{} instead extends the spins from
$t = -100\,M$ through merger and ringdown using post-Newtonian spin
evolution equations; these extended spins are explicitly unphysical, and
serve only as a convenient parametrization for constructing the
late-time fits~\cite{Varma:2019csw}. Late-time differences in these
auxiliary trajectories should therefore not be interpreted as errors in
physical horizon spins. Their effect on the waveform is nevertheless
included in the end-to-end mismatch.

\begin{figure*}[t]
  \centering
  \includegraphics[width=\textwidth]{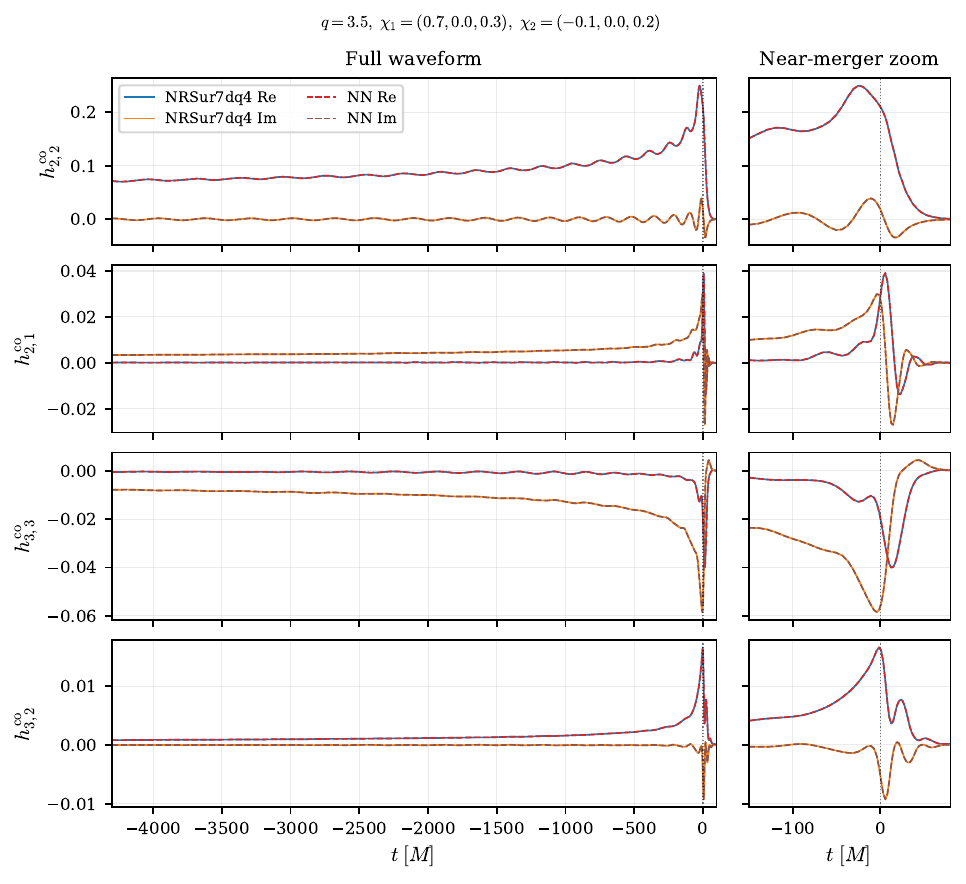}
  \caption{Real and imaginary parts of selected co-orbital modes
    $h_{\ell m}^\mathrm{coorb}$ for the binary in
    Fig.~\ref{fig:nn_waveform}. From top to bottom,
    the rows show the $(2,2)$, $(2,1)$, $(3,3)$, and $(3,2)$ modes. Blue
    solid and red dashed curves show the real parts from \nrsur{} and
    the neural surrogate, respectively. Orange solid and brown dashed
    curves show the corresponding imaginary parts. The left column
    spans the full waveform. The right column spans the interval from
    ${\sim}150\,M$ before to ${\sim}80\,M$ after merger, $t=0$
    (vertical dotted line).}
  \label{fig:nn_modes}
\end{figure*}

Finally, Fig.~\ref{fig:nn_modes} shows the real and imaginary parts of the
dominant co-orbital modes $(2,2)$, $(2,1)$, $(3,3)$, and $(3,2)$.
The \nrsur{} data are extracted directly from the gwsurrogate
\texttt{coorb\_sur} interface and compared with the corresponding NN
outputs on the native grid. These modes are reproduced with high
fidelity across the full waveform.

\subsection{Evaluation speed}
\label{sec:speed}

\begin{figure}[t]
  \centering
  \includegraphics[width=\columnwidth]{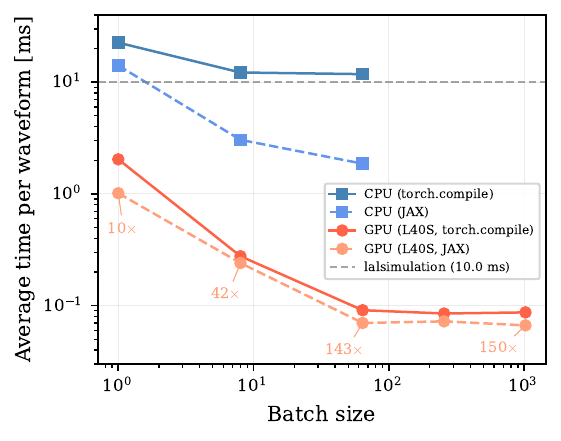}
  \caption{Average evaluation time per waveform as a function of the
    number of waveforms evaluated together. Each point is the elapsed
    time for a batch divided by its batch size. Curves show JAX and
    \texttt{torch.compile} on a CPU and on an NVIDIA L40S GPU.
    The benchmark uses $60\,\Msun$ waveforms with uniform time sampling
    at $4096\,\mathrm{Hz}$. 
    The horizontal dashed line marks the \textsc{lalsimulation}
    single-waveform reference ($10\,\mathrm{ms}$, single-threaded CPU).
    Speedup factors annotated on the JAX GPU curve are relative to
    \textsc{lalsimulation}.
  }
  \label{fig:latency}
\end{figure}

Figure~\ref{fig:latency} compares the JAX~\cite{jax2018github} and
PyTorch~\cite{paszke2019pytorch,ansel2024pytorch2} implementations on a
CPU and an NVIDIA L40S GPU. The PyTorch implementation uses
\texttt{torch.compile}. The batch size is the number of waveforms
evaluated together in one call. We report the average evaluation time
per waveform, obtained by dividing the elapsed time for a batch by the
batch size. For a batch of one, this is the usual evaluation latency,
while for larger batches its the average cost per waveform. 
The benchmark evaluates
$60\,\Msun$ waveforms at $4096\,\mathrm{Hz}$ and we compare
to the \textsc{lalsimulation} C implementation, which provides a single-threaded
CPU reference of $10\,\mathrm{ms}$ per waveform.

The plotted JAX values measure repeated calls after
JAX's Accelerated Linear Algebra (XLA) compiler~\cite{xla2017} has
built the executable. The first call for a new array shape, including a
new batch size, incurs a one-time compilation cost. For a batch of one,
this took ${\sim}13\,\mathrm{s}$ on the GPU and
${\sim}20\,\mathrm{s}$ on the CPU. Later calls with the same input shape
reuse the compiled function. The compilation time is not included in
Fig.~\ref{fig:latency}, which is justified as 
the startup cost is readily amortized in
parameter-estimation runs or other studies that evaluate many waveforms.

On the GPU, JAX is $10\times$ faster than \textsc{lalsimulation} for a
single waveform. The speedup increases to $42\times$ for a batch of 8
waveforms and $143\times$ for a batch of 64. At the largest tested batch
size of 1024, the speedup is $150\times$. JAX is also faster than
\texttt{torch.compile} at every batch size shown. We note that these timings use
NVIDIA's TensorFloat-32 (TF32) mode for eligible single-precision
matrix multiplications. Section~\ref{sec:implementations} discusses the
effect of TF32 on accuracy.

The average evaluation time per waveform initially falls rapidly as the
batch size increases, then changes little beyond a batch size of about 64.
Each call includes setup work that grows little with the batch size,
such as the overhead of launching GPU operations, and a larger batch
distributes this cost across more waveforms. Once the GPU reaches its
maximum sustained processing rate, increasing the batch size provides
little further benefit. Additional gains would require changes to the
calculation itself, such as smaller networks, more efficient matrix
multiplications, or lower-precision arithmetic.

To identify where the time is spent, we also timed each major operation
separately on an NVIDIA GH200 and report the median over repeated runs.
This breakdown uses a different GPU from Fig.~\ref{fig:latency}, so it
is intended to show relative costs rather than absolute latency. At a
batch size of 64, the data-piece networks account for ${\sim}45\%$ of
the sum of the separately timed operations. Most of this cost comes
from the $\Omega$ TCMLP, which processes a time-conditioned input at
each of the 230 native time samples. A double-precision cubic spline
resamples these values onto the uniform $4096\,\mathrm{Hz}$ physical-time
grid used in this benchmark, which contains 5327 samples. This
resampling contributes ${\sim}40\%$, and its cost grows with the number
of requested output samples. The Wigner-$D$ rotation contributes
${\sim}7\%$, while orbital-phase integration and strain projection each
contribute ${\sim}4\%$. Thus, the networks and output resampling account
for most of the sum of the separately timed operations.

At a batch size of 64, compiling the full pipeline as one operation
makes the end-to-end evaluation ${\sim}1.5\times$ faster than the sum
of these separately timed operations. XLA can combine operations and
avoid repeated GPU launch overhead in the full calculation. This
distinction is especially important for a batch size of 1, where launch
overhead dominates the separate timings and their sum substantially
overestimates the runtime of the compiled pipeline.

On the CPU, JAX takes ${\sim}14\,\mathrm{ms}$ for a batch of one,
which is about $40\%$ slower than the \textsc{lalsimulation} reference.
Batched evaluation reduces the average evaluation time to
${\sim}1.9\,\mathrm{ms}$ per waveform at a batch size of 64, giving a
speedup of approximately $5\times$.

\subsection{Implementation}
\label{sec:implementations}

\subsubsection{JAX implementation}
The \texttt{jax-nrsur7dq4-nn} package provides a JAX-native implementation
of the complete surrogate using the Equinox neural-network
library~\cite{kidger2021equinox}. It evaluates the dynamics and mode
networks, applies the Wigner-$D$ rotation, and assembles the strain.

The implementation supports both \texttt{jax.vmap} and
\texttt{jax.jit}. The former vectorizes the model over a batch of
parameters, while the latter compiles the vectorized calculation with
XLA. A single compiled call can therefore evaluate a large batch of
waveforms on a GPU, giving the speedups reported in
Sec.~\ref{sec:speed}.

The full pipeline is differentiable with respect to its model inputs.
JAX can differentiate scalar functions of the waveform with
\texttt{jax.grad} and construct waveform Jacobians with
\texttt{jax.jacfwd} without requiring hand-derived derivative routines.
The reference-frequency spin solver in Sec.~\ref{sec:fref} uses
\texttt{jax.jacfwd} to construct the Jacobian for each Newton iteration.

The physical-units interface is provided by
\texttt{NRSur7dq4NNModel}. It accepts the total mass in $M_\odot$ and
the luminosity distance in Mpc, then returns strain on a uniform
physical-time grid. Its conventions follow \textsc{gwsurrogate} and
\textsc{lalsimulation}.
The package is designed for GPU-accelerated parameter estimation and is
available at \url{https://github.com/mpuerrer/jax-nrsur7dq4-nn}.
The trained weights are distributed as a HDF5 file on Zenodo~\cite{nrsur7dq4nn_weights}.

\subsubsection{Numerical precision}
The data-piece networks and most tensor operations used for waveform
assembly run in single precision. On Ampere-class and newer NVIDIA GPUs,
XLA can use TensorFloat-32 (TF32) Tensor Cores for eligible batched
single-precision matrix multiplications. The batched GPU timings in
Sec.~\ref{sec:speed} use TF32 for these operations. At a batch size of
one, the networks instead use matrix-vector products that do not
generally engage the Tensor Cores.

We assess the effect of TF32 at the waveform level because its numerical
error depends on the calculation and data. Relative to a true
single-precision evaluation, TF32 perturbs the waveform itself by a flat
(white-noise) time-domain mismatch, the metric of
Sec.~\ref{sec:mismatch_defs}, of ${\sim}5\times10^{-6}$ (median
${\sim}3\times10^{-7}$), comparable to the surrogate's own time-domain
emulation error against \nrsur{} (Sec.~\ref{sec:e2e_accuracy}) and well
below its ${\sim}10^{-4}$ mismatch against NR.  The effect on the
reported statistics is smaller still: in a direct comparison of
500 waveforms on an L40S GPU, TF32 changed the sky-averaged
frequency-domain mismatch against \nrsur{} by ${\sim}10^{-7}$ in the
median and by at most a few${\times}10^{-6}$.  Individual waveforms shift
up or down at random rather than all in the same direction, so these
shifts largely cancel in the median and percentiles and do not affect
the distribution statistics reported in Sec.~\ref{sec:e2e_accuracy}. In
our timing tests, forcing strict IEEE single precision slowed evaluation
by a factor of ${\sim}2$ on the L40S and by up to ${\sim}8$ on the A100
without a meaningful improvement in waveform accuracy.

For numerically sensitive operations, the reported calculations enable
JAX's 64-bit mode. The orbital-phase integration, cubic-spline
interpolation onto the output grid, and construction of the Wigner-$D$
matrices use double precision. Complex quantities use complex128 where
needed. The orbital phase accumulates to several hundred radians and
enters the mode rotation through $e^{-im\phi_\mathrm{orb}}$. Lower
precision in the phase integration would therefore produce a coherent
phase error. Finally, in our tests, evaluating the $\Omega$ network in bfloat16 gave a further
${\sim}1.4\times$ end-to-end speedup, but produced a flat time-domain
mismatch of ${\sim}3\times10^{-4}$ against the single-precision
waveform. We therefore leave bfloat16 evaluation as an optional setting
in our public code.

\subsection{Spin specification at a reference frequency}
\label{sec:fref}

The \nrsur{} model natively
parameterize spins at the \emph{start} of the surrogate's internal time
grid ($t_0 \approx -4300\,M$).  Parameter-estimation (PE) codes,
however, conventionally specify the spin vectors
$\boldsymbol{\chi}_{1,2}$ at a user-chosen gravitational-wave reference
frequency $f_\mathrm{ref}$ (e.g.\ $20\,\mathrm{Hz}$).  For
non-precessing systems the spin directions are fixed, so this distinction
is irrelevant. For precessing binaries, however, the spin vectors
evolve substantially between the low-frequency reference and the start
of the surrogate's internal time grid.  We denote the bijective
spin-evolution map induced by the surrogate's spin dynamics by
$T:\theta_{t_0}\to\theta_{f_{\rm ref}}$; it carries the model's $t_0$
input spins to the physical spin vectors at $f_\mathrm{ref}$, so
specifying spins at $f_\mathrm{ref}$ and recovering the corresponding
$t_0$ model inputs amounts to evaluating the inverse $T^{-1}$.

In the \textsc{gwsurrogate} code, spins can also be specified at $f_\mathrm{ref}$ 
and are interpreted at the corresponding time $t_\mathrm{ref}$. The ODE dynamics
are initialized there and integrated both forward to merger and backward
to $t_0$.
This realizes the inverse map $T^{-1}$ (denoted $T_\mathrm{gws}^{-1}$
below) in a single ODE solve.  The native
\nrsur{} C implementation in \textsc{LALSuite} performs an analogous
procedure.

The neural surrogate's input spins, by contrast, are defined at $t_0$,
the start of its internal grid, so converting user-specified spins at
$f_\mathrm{ref}$ to model inputs requires solving the inverse problem.
Because the surrogate predicts the full spin trajectories
$\boldsymbol{\chi}_{1,2}(t)$ and the orbital frequency $\Omega(t)$ as
smooth, differentiable functions of the input parameters, we solve this
as a six-dimensional root-finding problem that realizes the neural
inverse $T_\mathrm{NN}^{-1}$.
Denoting the input spin components as the vector
$\boldsymbol{s} = (\chi_{1x}, \chi_{1y}, \chi_{1z},
\chi_{2x}, \chi_{2y}, \chi_{2z})$, we define the residual
\begin{equation}
  \mathbf{r}(\boldsymbol{s}) =
  \begin{pmatrix}
    \boldsymbol{\chi}_1\bigl(t_\mathrm{ref}(\boldsymbol{s})\bigr) \\[2pt]
    \boldsymbol{\chi}_2\bigl(t_\mathrm{ref}(\boldsymbol{s})\bigr)
  \end{pmatrix}
  -
  \begin{pmatrix}
    \boldsymbol{\chi}_1^\mathrm{target} \\[2pt]
    \boldsymbol{\chi}_2^\mathrm{target}
  \end{pmatrix} ,
\label{eq:fref_residual}
\end{equation}
where $t_\mathrm{ref}(\boldsymbol{s})$ is the time at which
$\Omega(t) = \pi f_\mathrm{ref}$ for the current trial parameters.
The coupling between $\boldsymbol{s}$ and $t_\mathrm{ref}$ arises
because the orbital frequency is conditioned on the predicted spin
trajectories (Sec.~\ref{sec:nrsur_structure}).

We solve $\mathbf{r}(\boldsymbol{s}) = 0$ with Newton's method,
\begin{equation}
  \boldsymbol{s}_{k+1} = \boldsymbol{s}_k
    - \mathbf{J}^{-1}(\boldsymbol{s}_k)\,\mathbf{r}(\boldsymbol{s}_k) \,,
\label{eq:newton}
\end{equation}
where the $6 \times 6$ Jacobian
$J_{ij} = \partial r_i / \partial s_j$ is computed via JAX automatic
differentiation at negligible cost.  The initial guess is the identity
mapping $\boldsymbol{s}_0 = (\boldsymbol{\chi}_1^\mathrm{target},\,
\boldsymbol{\chi}_2^\mathrm{target})$.  At each iteration only the
neural network forward pass for the dynamics models (spin trajectories
and $\Omega$) is evaluated while the more expensive mode prediction,
Wigner-$D$ rotation, and strain assembly are skipped.  The solver
typically converges in fewer than ten iterations.
Finally, to avoid the post-merger regime where $\Omega(t)$ oscillates and the
spline inversion becomes unreliable, the allowed reference frequency is
capped at $80\%$ of the peak orbital frequency.  This still covers the
entire inspiral band relevant for PE.

\begin{figure*}[t]
  \centering
  \includegraphics[width=\textwidth]{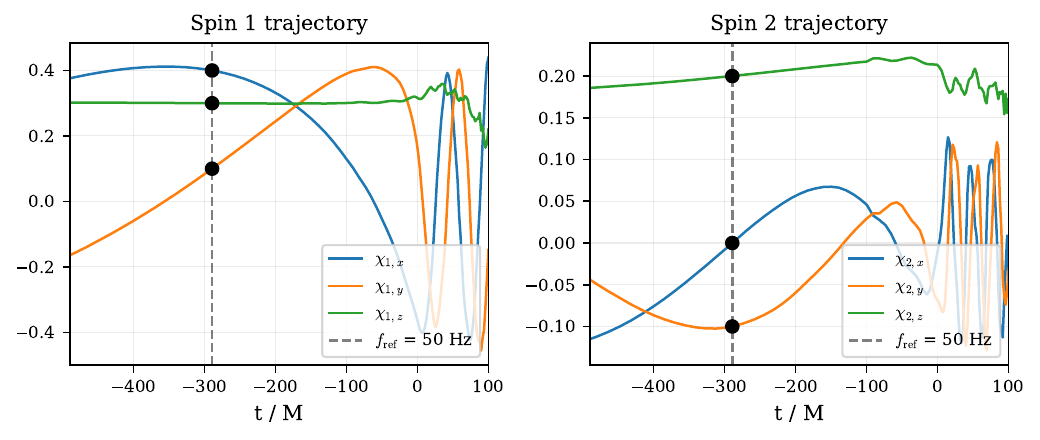}
  \caption{Spin trajectories predicted by the neural surrogate for the
    heavier (left) and lighter (right) black holes in a precessing binary
    with $q = 3$ and $M = 60\,\Msun$. The model input spins have been
    determined by the reference-frequency solver so that the trajectories
    match the user-requested spin vectors (black dots) at
    $f_\mathrm{ref} = 50\,\mathrm{Hz}$ (dashed line). This agreement
    illustrates how the solver converts spins specified at
    $f_\mathrm{ref}$ into the corresponding model inputs at $t_0$.
    See Sec.~\ref{sec:fref} for details of the inversion procedure and
    its accuracy.}
  \label{fig:spins_at_fref}
\end{figure*}

Figure~\ref{fig:spins_at_fref} illustrates an example.  The user
requests $\boldsymbol{\chi}_1 = (0.4,\, 0.1,\, 0.3)$ and
$\boldsymbol{\chi}_2 = (0.0,\, -0.1,\, 0.2)$ at
$f_\mathrm{ref} = 50\,\mathrm{Hz}$ for a binary with
$q = 3$ and $M = 60\,\Msun$.
The solver finds the model input spins whose predicted
trajectories pass through the requested values (black dots) at
the corresponding reference time (dashed vertical line).

Because the $f_\mathrm{ref}\!\to\!t_0$ inversion of Eq.~\eqref{eq:newton}
is driven by the network-predicted spin trajectories
$\boldsymbol{\chi}_{1,2}(t)$, any error in the spin-dynamics networks
propagates into the recovered $t_0$ input spins, and hence into
parameter estimation whenever spins are specified at $f_\mathrm{ref}$.
We quantify this with a controlled sensitivity test.  Drawing $5000$
precessing configurations from the canonical isotropic-direction,
magnitude-uniform spin prior ($q\in[1,4]$, $|\boldsymbol{\chi}_i|\le0.8$,
$M\in[60,120]\,\Msun$, $f_\mathrm{ref}=20\,\mathrm{Hz}$), we map each
back to $t_0$ in two ways: (i) with the neural solver $T_\mathrm{NN}^{-1}$
of Eq.~\eqref{eq:newton} and (ii) with the exact \nrsur{} spin ODE integrated
backward by \textsc{gwsurrogate}, $T_\mathrm{gws}^{-1}$. We then generate
two neural-surrogate waveforms that differ \emph{only} in their $t_0$
spin input.  The resulting mismatch isolates the spin-network
contribution to the inversion, independent of the rest of the waveform
pipeline.  It is a self-consistency test rather than a measure of
absolute accuracy, since a bias common to $T_\mathrm{NN}$ and
$T_\mathrm{gws}$ would cancel; absolute accuracy is bounded separately
by the end-to-end mismatch of Sec.~\ref{sec:e2e_accuracy}.

The recovered $t_0$ input spins differ between the two maps by a median
of $8\times10^{-4}$ and $1.5\times10^{-3}$ (Cartesian magnitude) for the
primary and secondary spin, respectively, and the corresponding
aLIGO-PSD-weighted mismatch between the two waveforms has a median of
$9.5\times10^{-7}$ and a 95th percentile of $4\times10^{-5}$.  The error
is strongly concentrated in the highly precessing regime
(Table~\ref{tab:spin_inversion}), where the median mismatch grows by an order
of magnitude from $\chip\lesssim0.1$ to $\chip\gtrsim0.5$.
Expressed as an indistinguishability criterion~\cite{Lindblom:2008cm}, the two
inversions satisfy $M<1/(2\rho^2)$ for $99.2\%$, $98.0\%$, and $95.4\%$
of the prior at network SNR $\rho=20$, $50$, and $100$, respectively.
We conclude that the choice of spin-evolution code is irrelevant for essentially all
current detections and matters only for the loudest, most strongly
precessing signals.  These fractions are computed from the $99\%$ of draws for which both
inversions converge. The remaining ${\sim}1\%$ are strongly precessing,
with median $\chip=0.63$ compared with $0.41$ for the prior as a whole.
Their exclusion makes the quoted fractions mildly optimistic.

\begin{table}[t]
  \centering
  \caption{Spin-network inversion error measured using the
aLIGO-PSD-weighted mismatch. We compare neural-surrogate waveforms whose
$t_0$ input spins are recovered either with the neural solver
$T_\mathrm{NN}^{-1}$ or by backward integration of the \nrsur{} spin
ODE with \textsc{gwsurrogate} ($T_\mathrm{gws}^{-1}$). Results are
grouped by $\chip$ at $f_\mathrm{ref}=20\,\mathrm{Hz}$ and use
$5000$ draws with $q\in[1,4]$, $|\boldsymbol{\chi}_i|\le0.8$, and
$M\in[60,120]\,\Msun$.}
  \label{tab:spin_inversion}
  \begin{ruledtabular}
  \begin{tabular}{lcccc}
    $\chip$ & $N$ & median $M$ & 95th $M$ & max $M$ \\
    \hline
    $[0.0, 0.1)$ &  $202$ & $3.3\times10^{-7}$ & $3.2\times10^{-6}$ & $1.5\times10^{-5}$ \\
    $[0.1, 0.3)$ & $1271$ & $3.4\times10^{-7}$ & $5.8\times10^{-6}$ & $5.2\times10^{-4}$ \\
    $[0.3, 0.5)$ & $1824$ & $7.2\times10^{-7}$ & $1.4\times10^{-5}$ & $4.2\times10^{-3}$ \\
    $[0.5, 0.8)$ & $1654$ & $3.3\times10^{-6}$ & $2.4\times10^{-4}$ & $1.5\times10^{-2}$ \\
  \end{tabular}
  \end{ruledtabular}
\end{table}

\section{Applications}

\subsection{Parameter estimation: GW150914}
\label{sec:pe}

As a concrete demonstration, we have integrated the neural surrogate
into a standard \textsc{bilby}~\cite{Ashton2019}\,+\,\textsc{dynesty}
~\cite{2020MNRAS.493.3132S,sergey_koposov_2025_17268284}
nested-sampling workflow and accelerated it on a single NVIDIA GH200
Grace-Hopper module.  The central difficulty in exploiting a GPU within
\textsc{dynesty} is that the sampler advances through a pool of worker
processes, each of which requests only a \emph{single} likelihood
evaluation at a time.  Dispatched individually, these calls incur the
batch-size one latency of Fig.~\ref{fig:latency}, where per-call dispatch
overhead dominates and the GPU's batched-throughput advantage has not
yet set in. Consequently, a naive \textsc{bilby}\,+\,\textsc{dynesty} run with the JAX
model is several times slower than the CPU waveform and
leaves the GPU almost idle. %
We instead route every worker's request through a lightweight
inter-process queue to a single resident GPU server process.  The
server collects the requests that arrive within a short time window (or
up to a maximum batch size), stacks them into one parameter array,
evaluates the batched, \texttt{jax.vmap}-ed and JIT-compiled likelihood
in a single GPU call, and returns each scalar to its originating worker.
Because \textsc{dynesty}'s pool naturally keeps many proposals in flight,
this opportunistic batching reaches an average of ${\sim}28$ waveforms
per GPU call over a full run, with a sustained throughput of
${\sim}2.6\times10^{3}$ likelihood evaluations per second for an analysis
of GW150914. This places the likelihood in the favorable region of the
batch-size scaling curve (Fig.~\ref{fig:latency}) at no cost to the sampler,
which is unmodified. The GH200's unified Grace-Hopper memory eliminates
host--device transfer overhead for the per-call parameter and strain arrays.

We sample with the spin vectors defined at the surrogate's native start
time $t_0$ (Sec.~\ref{sec:fref}) rather than at a reference frequency.
This avoids the per-likelihood Newton solve of the
$f_\mathrm{ref}\!\to\!t_0$ inversion inside the sampling loop, where it
would be incurred ${\sim}10^{7}$ times.  The resulting posterior is
mapped to the conventional $f_\mathrm{ref}$ spin convention in a single
post-processing pass over the ${\sim}10^{4}$ posterior samples, using
the forward spin evolution of either the neural surrogate or
\nrsur{}. The remapping leaves the physical waveform, and therefore the
matched-filter likelihood, unchanged.

Mapping posterior samples from $t_0$ to $f_\mathrm{ref}$ also
transforms the spin prior. Compared with imposing the prior directly at
$f_\mathrm{ref}$, the mapped prior preserves the spin-magnitude and
overall-orientation distributions but slightly reshapes the relative
spin-tilt distribution through precession. Appendix~\ref{app:prior_map}
shows that this difference is negligible for the weakly precessing
GW150914. We restrict all samples to the \nrsur{} calibration region,
$1\leq q\leq4$ and $|\boldsymbol{\chi}_i|\leq0.8$, so no waveform
requires extrapolation.

Table~\ref{tab:pe_gw150914} compares three analyses of GW150914 using
the same priors and data. The GPU-batched neural surrogate and \nrsur{}
through \textsc{LALSuite}'s time-domain interface (LAL-TD) use the same
signal conditioning. The third analysis evaluates \nrsur{} with the
standard \textsc{bilby} frequency-domain generator (LAL-FD), which uses
\textsc{LALSuite}'s internal conditioning. The neural and LAL-TD
pipelines apply a Tukey window with $\alpha=0.1$ to the fixed-length
time-domain waveform, followed by an FFT and an epoch correction. They
use neither a high-pass filter nor zero-padding. LAL-FD instead applies
\textsc{SimInspiralFD} conditioning, including a chirp-time-dependent
start taper, a Butterworth high-pass filter at $f_\mathrm{low}$, and
zero-padding to a power-of-two length. The resulting waveforms differ
only near $f_\mathrm{low}$, producing the conditioning systematic
discussed below. The fixed grid and FFT size are required by JAX's
static-shape constraint and allow the full likelihood to remain
differentiable, \texttt{jax.jit}-compiled, and
\texttt{jax.vmap}-batched. By contrast, \textsc{SimInspiralFD} chooses
a parameter-dependent transform length and relies on
\textsc{LALSuite} C code that cannot be differentiated or batched on a
GPU.

All three analyses use \textsc{dynesty} with a pool of $60$ worker
processes. The two \textsc{LALSuite} runs were each allocated $64$ cores
of an AMD EPYC node, either an EPYC~7H12 at $2.60\,\mathrm{GHz}$ or an
EPYC~7763 at $2.45\,\mathrm{GHz}$ base clock, so their runtimes may
differ modestly with processor generation and operating frequency; the
neural-surrogate run used the same pool size, with every worker's
likelihood request served by the single resident GH200 process
described above.

The neural-surrogate analysis completes in under four hours. This is
more than twice as fast as LAL-TD and roughly three times faster than
LAL-FD. All three analyses use the same noise evidence. As an
end-to-end comparison of their likelihood surfaces, we report the
signal-to-noise log Bayes factors in Table~\ref{tab:pe_gw150914}.
The Bayesian evidence averages the likelihood over the prior, and its
natural logarithm is measured in units of nats. The three values span
${\sim}1$~nat. The neural-surrogate evidence is ${\sim}1$~nat below both
\textsc{LALSuite} results. Relative to the reported nested-sampling
uncertainties, this is a formal $3$--$4\sigma$ difference, accompanied
by a $0.3$-nat reduction in the largest sampled log-likelihood. This is
consistent with its ${\sim}6\times10^{-4}$ effective mismatch against
\nrsur{} (Sec.~\ref{sec:e2e_accuracy}).

\begin{figure*}[!tp]
  \centering
  \includegraphics[width=\textwidth]{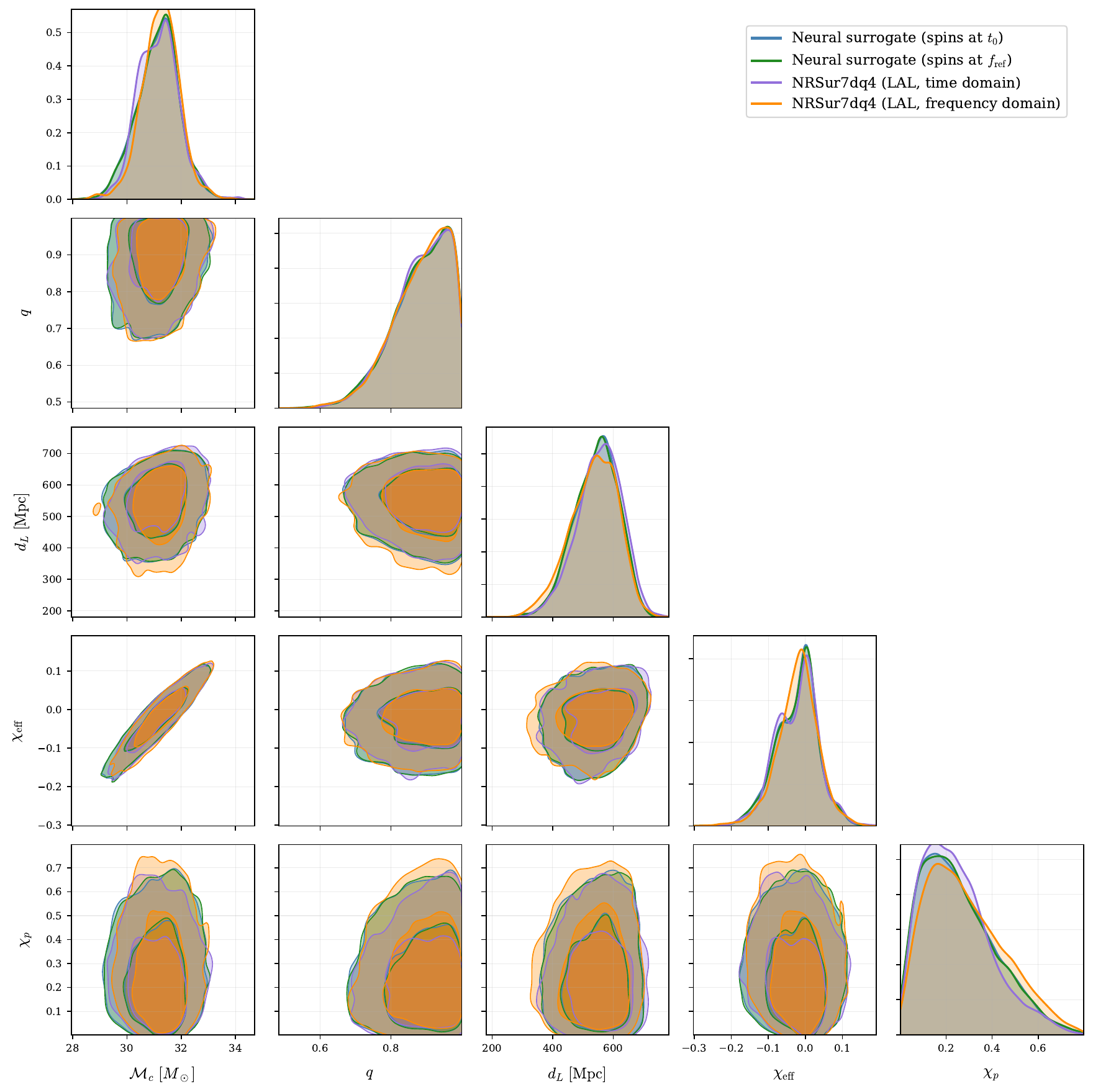}
  \caption{
    Marginalized posterior distributions for GW150914 from four
    \nrsur{}-based analyses with identical priors and data settings.
    The two neural-surrogate results show the posterior in its native
    $t_0$ spin convention (\emph{Neural surrogate, spins at $t_0$}) and
    after remapping to $f_\mathrm{ref}=20\,\mathrm{Hz}$ with the
    \textsc{gwsurrogate} forward spin evolution
    (\emph{Neural surrogate, spins at $f_\mathrm{ref}$}). The two
    reference analyses evaluate \nrsur{} through
    \textsc{LALSuite}'s time-domain interface with matched conditioning
    (\emph{NRSur7dq4, LAL, time domain}) and through the standard
    \textsc{bilby} frequency-domain generator
    (\emph{NRSur7dq4, LAL, frequency domain}). Shown are the chirp mass
    $\mathcal{M}_c$, mass ratio $q$, luminosity distance $d_L$, and the
    effective and precessing spin parameters $\chieff$ and $\chip$.
    Diagonal panels show the one-dimensional marginals. Off-diagonal
    panels show the $68\%$ and $95\%$ two-dimensional credible regions.
    All four analyses are mutually consistent.}
  \label{fig:pe_gw150914}
\end{figure*}

Figure~\ref{fig:pe_gw150914} compares the recovered posteriors.  The
neural-surrogate result is shown both in its native $t_0$ spin
convention and after the post-processing remap to $f_\mathrm{ref}=20\,
\mathrm{Hz}$ (here using the \textsc{gwsurrogate} forward spin
evolution), alongside the two \textsc{LALSuite} \nrsur{} analyses.  The
marginalized posteriors for chirp mass, mass ratio, luminosity distance,
and the effective and precessing spin parameters
$(\chieff,\,\chip)$ are mutually consistent across all four analyses.
We find that the neural surrogate reproduces the \nrsur{} parameter estimates, and
the $t_0$ and remapped spin posteriors agree, confirming that sampling
in the native convention with a post-hoc spin evolution introduces no
bias in the inferred parameters.  

We quantify these statements with the
Jensen--Shannon (JS) divergence between the marginal
posteriors~\cite{Romero-Shaw:2020owr}, the standard measure of posterior
agreement in gravitational-wave parameter estimation.  The $t_0$ and
remapped neural-surrogate posteriors agree to $\le2\times10^{-3}\,
\mathrm{bit}$ across all five parameters, at or below the
${\sim}3\times10^{-3}\,\mathrm{bit}$ same-posterior sampling-noise floor at
these chain lengths (${\sim}5\times10^{3}$ effective samples) and below the
${\sim}2\times10^{-3}\,\mathrm{bit}$ value conventionally taken as
indistinguishable, so for GW150914 the post-hoc spin evolution leaves no
detectable imprint.  The neural surrogate and the two \textsc{LALSuite} \nrsur{}
pipelines differ by at most $1.5\times10^{-2}\,\mathrm{bit}$, comparable to
and below the $1.7\times10^{-2}\,\mathrm{bit}$ that separates the two
\textsc{LALSuite} pipelines (time- versus frequency-domain conditioning)
from \emph{each other}. That is the neural surrogate reproduces \nrsur{} to within
the waveform-conditioning systematic that already distinguishes the two
reference pipelines.

\begin{table}[t]
  \centering
  \caption{Parameter estimation on GW150914 with \nrsur{} waveforms,
    \textsc{dynesty} nested sampling ($n_\mathrm{live}=1000$,
    acceptance-walk).  \emph{NN (GPU)}: neural surrogate with the
    GPU-batched likelihood on a single GH200.  \emph{LAL-TD}: \nrsur{}
    via \textsc{LALSuite}'s time-domain interface with conditioning
    matched to the neural-surrogate likelihood.  \emph{LAL-FD}: \nrsur{}
    via the standard \textsc{bilby} frequency-domain waveform generator,
    which applies \textsc{LALSuite}'s own internal signal conditioning.
    $\ln\mathcal{L}_\mathrm{r}$ is the log-likelihood ratio relative to
    the noise hypothesis (maximized over the posterior samples) and
    $\mathcal{B}_\mathrm{s/n}$ the signal-versus-noise Bayes factor
    ($\mathcal{Z}_\mathrm{signal}/\mathcal{Z}_\mathrm{noise}$).  All
    quantities are read from the saved \textsc{bilby} result files.
    }
  \label{tab:pe_gw150914}
  \begin{ruledtabular}
  \begin{tabular}{lccc}
    & NN (GPU) & LAL-TD & LAL-FD \\
    \hline
    Wall time [h]                     & $3.7$  & $8.1$  & $12.0$ \\
    Iterations                        & $33\,912$ & $33\,202$ & $32\,934$ \\
    Likelihood calls                  & $3.5\times10^{7}$ & $3.6\times10^{7}$ & $3.5\times10^{7}$ \\
    $\max \ln \mathcal{L}_\mathrm{r}$ & $279.34$ & $279.62$ & $279.63$ \\
    $\ln \mathcal{B}_\mathrm{s/n}$    & $248.84 \pm 0.19$ & $249.72 \pm 0.19$ & $250.01 \pm 0.20$ \\
  \end{tabular}
  \end{ruledtabular}
\end{table}

\subsection{Injection recovery for a strongly precessing source}
\label{sec:pe_injection}

The GW150914 analysis demonstrates faithful recovery on real data
for a weakly precessing source.  To verify that the neural surrogate
returns unbiased parameter estimates in the regime where precession is
strong, and to do so against a known ground truth, we additionally
analyse a simulated signal injected into Gaussian noise coloured by the
detector design power spectral densities.  The injected binary has
component masses $m_1=62.5\,M_\odot$ and $m_2=25\,M_\odot$ (mass ratio
$q=0.4$, chirp mass $\mathcal{M}_c=33.7\,M_\odot$) and large,
misaligned spins ($a_1=0.7$ tilted at $60^\circ$, $a_2=0.5$ tilted at
$29^\circ$), giving a strong in-plane spin component
($\chip=0.61$) alongside a moderate aligned component
($\chieff=0.38$).  The source is placed at $d_L=600\,\mathrm{Mpc}$ with
inclination $\theta_{JN}=60^\circ$, yielding a network signal-to-noise
ratio of $33.9$ ($26.8$ in H1, $20.8$ in L1).  The signal is injected
with \nrsur{} through \textsc{LALSuite}'s time-domain interface and
recovered with the same three pipelines used above: the GPU-batched
neural surrogate, \nrsur{} via \textsc{LALSuite} time domain (matched
conditioning), and \nrsur{} via the \textsc{bilby} frequency-domain
generator. We use identical broad priors restricted to the \nrsur{}
calibration region ($q\ge1/4$, spin magnitudes $\le0.8$).  We use
$n_\mathrm{live}=2000$ live points.

\begin{figure*}[!tp]
  \centering
  \includegraphics[width=\textwidth]{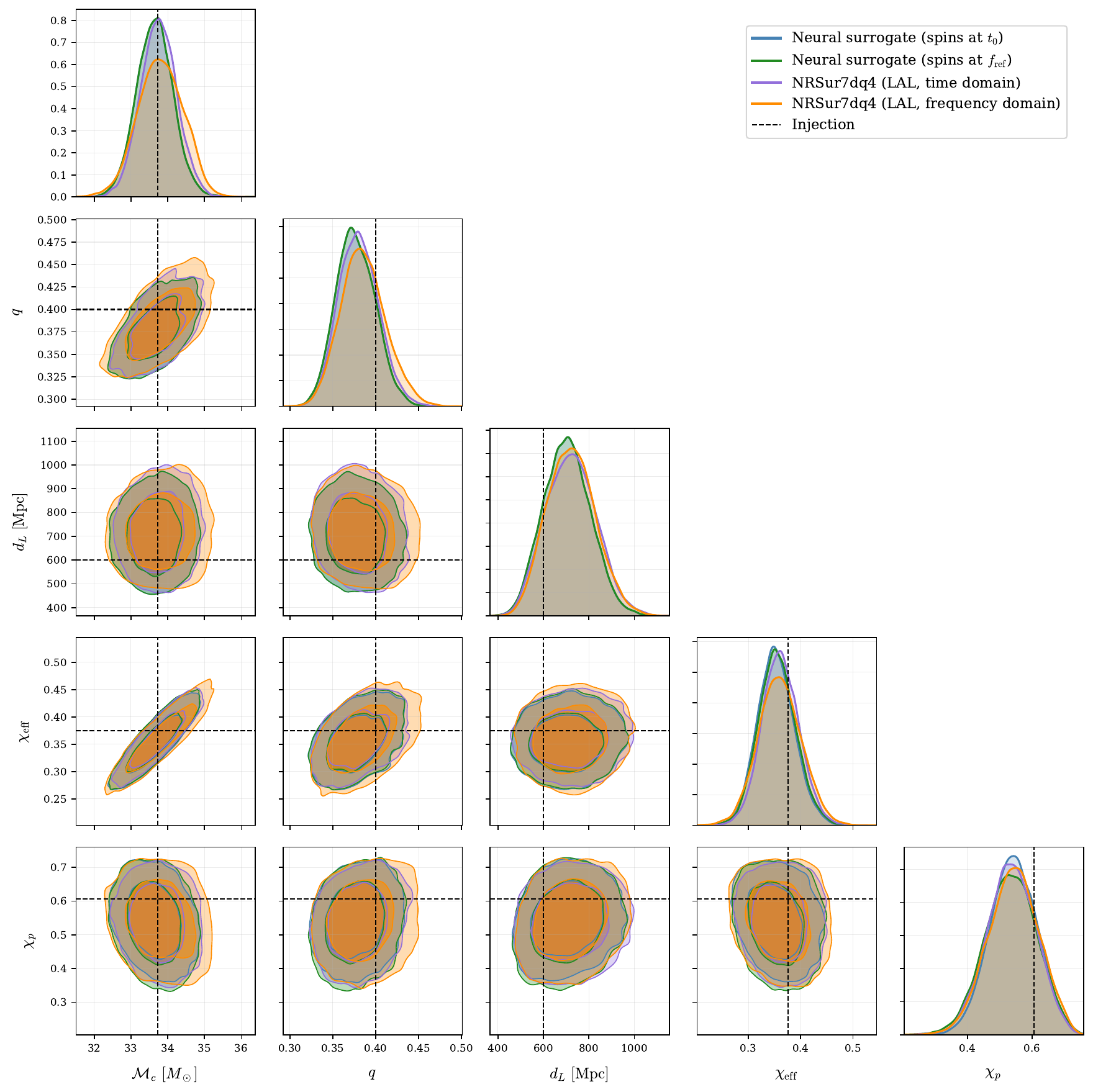}
  \caption{Marginalized posterior distributions for the strongly precessing
injection of Sec.~\ref{sec:pe_injection}. The comparison includes the
neural surrogate sampled in its native $t_0$ spin convention
(\emph{NNSur, spins at $t_0$}), the same posterior remapped to
$f_\mathrm{ref}=20\,\mathrm{Hz}$ using the \textsc{gwsurrogate}
forward spin evolution
(\emph{NNSur, spins at $f_\mathrm{ref}$}), and \nrsur{} evaluated
through the \textsc{LALSuite} time- and frequency-domain interfaces
(\emph{NRSur7dq4, LAL TD/FD}). Shown are the chirp mass
$\mathcal{M}_c$, mass ratio $q$, luminosity distance $d_L$, and the
effective and precessing spin parameters $\chieff$ and $\chip$. Dashed
lines mark the injected values. Diagonal panels show the
one-dimensional marginals. Off-diagonal panels show the $68\%$ and
$95\%$ two-dimensional credible regions. All analyses recover the
injection and are mutually consistent.}
  \label{fig:pe_injection}
\end{figure*}

Figure~\ref{fig:pe_injection} shows the recovered posteriors, and
Table~\ref{tab:pe_injection} lists the injected and recovered values.
All three pipelines recover every parameter within its $90\%$ credible
interval, and the neural-surrogate medians track the \nrsur{} medians
closely.  The luminosity distance is recovered slightly high
(median ${\sim}700\,\mathrm{Mpc}$ against the injected $600\,\mathrm{Mpc}$)
by \emph{all three} pipelines alike, a consequence of the
distance--inclination degeneracy and the particular noise
realization, not of the surrogate, since the \textsc{LALSuite}
\nrsur{} analyses show the same offset.  As for GW150914, the
neural-surrogate posterior sampled in the native $t_0$ convention and
the same posterior remapped to $f_\mathrm{ref}=20\,\mathrm{Hz}$ (using
the \textsc{gwsurrogate} forward spin evolution) are visually
indistinguishable.  We use the
\textsc{gwsurrogate} spin evolution rather than the surrogate's own
spin-trajectory data pieces. We find that for large in-plane spins (ie strong precession) 
of this source the latter are not yet accurate enough to integrate the
precession dynamics back to $f_\mathrm{ref}$ without a visible bias in
the inferred $\chieff$ and $\chip$. We leave imporoves to the spin networks
as future work. 

We quantify these agreements with the JS divergence between the
marginal posteriors.  The $t_0$ and remapped neural-surrogate
posteriors agree to better than $6\times10^{-3}\,\mathrm{bit}$ across all
five parameters, 
confirming that the post-hoc spin remap remains unbiased even when
precession is strong.  The neural surrogate and the matched
time-domain \nrsur{} pipeline differ by at most
$1.1\times10^{-2}\,\mathrm{bit}$ (in $\chieff$), which is \emph{below} the
$1.9\times10^{-2}\,\mathrm{bit}$ that separates the two \textsc{LALSuite}
pipelines from each other through their differing signal conditioning.
That is, the neural surrogate reproduces \nrsur{} to within the
waveform-conditioning systematic already present between the reference
pipelines.  The log-evidences reinforce this picture.  All three
analyses share the same noise evidence; the neural surrogate
($\ln\mathcal{B}_\mathrm{s/n}=530.6$) agrees with the matched
time-domain \nrsur{} run ($530.9$) to $0.3$~nat, while the
frequency-domain pipeline ($528.2$) lies $2.7$~nat lower. The
neural-surrogate evidence is closer to the matched \nrsur{} reference
than the two \textsc{LALSuite} pipelines are to each other.

The neural-surrogate analysis completes in $14.6$~h of wall time on
a single GH200 (worker pool of $64$), against $36.5$~h and $38.9$~h for
the time- and frequency-domain \textsc{LALSuite} runs on the CPU nodes
described in Sec.~\ref{sec:pe} (pool of $60$), a ${\sim}2.5\times$ speedup on
a problem substantially more demanding than GW150914 (twice the live
points, higher SNR, and strong precession).  Together with the
GW150914 result, this establishes that the GPU-accelerated neural
surrogate delivers \nrsur{}-faithful, unbiased parameter estimates for
precessing binaries at a fraction of the cost of the native
implementation.

\begin{table}[t]
  \centering
  \caption{Injected and recovered parameters for the strongly precessing injection
of Sec.~\ref{sec:pe_injection}. Recovered values are reported as the
posterior median with the symmetric $90\%$ credible interval.
\emph{NN (GPU)} denotes the neural surrogate sampled in the $t_0$ spin
convention. \emph{LAL-TD} and \emph{LAL-FD} denote \nrsur{} evaluated
through the time- and frequency-domain \textsc{LALSuite} interfaces,
respectively. The injected values of $\mathcal{M}_c$, $\chieff$, and
$\chip$ are derived from the component masses, spin magnitudes, and
tilts.}
  \label{tab:pe_injection}
  \begin{ruledtabular}
  \begin{tabular}{lcccc}
    & Injected & NN (GPU) & LAL-TD & LAL-FD \\
    \hline
    \rule{0pt}{2.6ex}%
    $\mathcal{M}_c\,[M_\odot]$ & $33.72$ & $33.65^{+0.81}_{-0.81}$ & $33.75^{+0.82}_{-0.80}$ & $33.80^{+0.98}_{-1.00}$ \\[2pt]
    $q$                        & $0.400$ & $0.376^{+0.039}_{-0.034}$ & $0.379^{+0.040}_{-0.036}$ & $0.384^{+0.047}_{-0.039}$ \\[2pt]
    $d_L\,[\mathrm{Mpc}]$      & $600$   & $703^{+172}_{-163}$ & $721^{+181}_{-176}$ & $721^{+183}_{-165}$ \\[2pt]
    $\chieff$                  & $0.375$ & $0.352^{+0.061}_{-0.054}$ & $0.361^{+0.061}_{-0.056}$ & $0.359^{+0.069}_{-0.066}$ \\[2pt]
    $\chip$                    & $0.606$ & $0.543^{+0.118}_{-0.119}$ & $0.536^{+0.120}_{-0.128}$ & $0.544^{+0.124}_{-0.133}$ \\
  \end{tabular}
  \end{ruledtabular}
\end{table}

\section{Discussion and conclusions}
\label{sec:discussion}

We have built a fast, accurate, and differentiable neural-network
surrogate for \nrsur{}, itself a surrogate of numerical-relativity
(NR) simulations, covering generic precessing binary black holes with
mass ratios up to $4$ and dimensionless spin magnitudes up to $0.8$.
Its emulation error is comfortably subdominant to the error of
\nrsur{} relative to NR, and batched GPU evaluation is up to two orders
of magnitude faster than the existing \textsc{lalsimulation}
implementation. To our knowledge, this is the first precessing
neural-network NR surrogate to combine NR-faithful accuracy with a fully
differentiable, GPU-accelerated waveform-to-likelihood pipeline. These
capabilities enable gradient-based and GPU-accelerated inference and
offer a path from current multi-day parameter-estimation runs toward
sub-hour analyses of precessing sources. We first discuss the model
construction and its main methodological lessons, then place its
accuracy, speed, and inference applications in context. We close with
the model's limitations and broader directions for future work.

\subsection{Construction and modeling choices}

Beyond the model itself, two findings emerged that we believe transfer to waveform surrogate modeling more broadly. First, \emph{the quality of the training target can matter more than the choice of architecture}: smoothing the orbital-frequency target in the post-merger region (Sec.~\ref{sec:smoothing}) sharply reduces its variance and improves end-to-end accuracy more than any architectural change we explored, whereas an unsmoothed target limits the achievable accuracy regardless of network size. Second, \emph{network capacity is best allocated by an explicit error budget rather than spread uniformly across data pieces}: the end-to-end mismatch is dominated by a few data pieces, and directing capacity to where it most reduces the predicted mismatch (Secs.~\ref{sec:error_decomp_method},~\ref{sec:network_sizing}) yields a mixed bank of architectures rather than a one-size-fits-all network.

The model follows \nrsur{}'s piecewise structure
(Sec.~\ref{sec:decomposition}, Fig.~\ref{fig:pipeline}) but replaces its
dynamics ODE with direct regression of the orbital frequency,
co-precessing quaternion, and spin trajectories.  Most data pieces use
feedforward MLPs with GELU activations. The orbital frequency, however,
uses a time-conditioned MLP with Fourier feature encoding and
instantaneous spin inputs (Sec.~\ref{sec:tcmlp}), mirroring the
dependence of \nrsur{}'s ODE right-hand side on the instantaneous spin
state. Network sizes are chosen systematically rather than by hand. A
leave-one-out decomposition and Monte Carlo perturbation analysis
(Sec.~\ref{sec:error_decomp_method}) identify which modeled quantities
contribute most strongly to waveform error, and a greedy allocation
(Sec.~\ref{sec:network_sizing}) directs the fixed parameter budget to
the networks where additional capacity most reduces the predicted
mismatch.

Neural-network surrogate waveform models for binary black holes
have progressed from early demonstrations and refinements of neural
interpolation over reduced waveform
representations~\cite{Chua:2018woh,Nousi:2021arn,Fragkouli:2022lpt},
through aligned-spin inspiral-merger-ringdown and higher-mode
models~\cite{Khan:2020fso,Grimbergen:2024kwk}, to NR-informed
nonprecessing surrogates with parameter-estimation
validation~\cite{GramaxoFreitas:2024bpk,Theodoropoulos:2026wkj} and a
precessing higher-mode EOB emulator with restricted spin
freedom~\cite{Thomas:2022rmc}. Concurrent work by Whittall \&
Pratten~\cite{Whittall:2026yvp} largely removes the earlier coverage
and validation restrictions. Their model spans the full
seven-dimensional quasicircular intrinsic parameter space within
$q\leq10$ and is validated in Bayesian analyses of both injections and
observed events. It nevertheless models only six positive-$m$
co-precessing-frame modes and reconstructs the negative-$m$ modes using
approximate conjugate symmetry. Their current implementation therefore
omits $m=0$ modes and the co-precessing-frame mode asymmetries retained
by \nrsur{}.

Our work is most similar to that of Whittall \& Pratten,
and we briefly summarize some differences here.
Whittall \& Pratten emulate 
\textsc{SEOBNRv5PHM} over a broader parameter domain.
We emulate the NR-trained \nrsur{} model over its
calibration region and retain all 21 modes with $\ell\leq4$. 
Their construction uses reduced
bases and empirical interpolation and predicts the orbital phase
directly. Ours predicts the sampled quantities directly, including the
orbital frequency and spin trajectories, and obtains the phase by
integration. Our construction also uses error-budget-driven network
sizing, provides a learned map for spins specified at a reference
frequency, and implements a differentiable JAX waveform-to-likelihood
pipeline. Their model also exhibits a known accuracy degradation when
a quaternion component vanishes at the waveform start time. This
singularity arises from their reconditioning strategy and is absent
from our approach.

For our orbital-frequency target, we also tested the
reduced-basis/EIM representation of the type used by Whittall \&
Pratten. Holding the training set and PieceMLP
hidden-layer architecture fixed, we find that direct per-timestep prediction
produced a lower orbital-frequency MSE than the reduced-basis models
(Sec.~\ref{sec:architecture}). This result is specific to our setup and
is not intended as a general comparison of the two methodologies. At
the chosen compression level, even the exact reduced-basis projection
gave a larger inspiral error than the direct model. Standard greedy EIM
selection also placed more than half of its nodes at $t\geq40M$, in the
harder-to-model post-merger and ringdown region. We have not tested
whether these findings extend to other data pieces or parameter spaces,
or whether weighted or constrained node selection would change the
result.

\subsection{Accuracy and speed}

Over the mass range common to Fig.~\ref{fig:fd_vs_mass} and Fig.~4 of
Varma et al.~\cite{Varma:2019csw}, our mismatches against \nrsur{} have
medians several times lower and 95th percentiles about an order of
magnitude lower than the reported \nrsur{} mismatches against NR. The
validation samples and mismatch summaries differ, so this comparison is
only indicative. The separation nevertheless suggests that the NN
emulation error is smaller than the error of \nrsur{} against NR.
Figures~\ref{fig:nn_waveform}--\ref{fig:nn_modes} illustrate this
agreement for the representative high-mass-ratio, high-spin example of
Sec.~\ref{sec:waveform_examples}. For example, at $M=60\,M_\odot$, our surrogate has a median
sky-averaged mismatch of $8 \times 10^{-5}$ against \nrsur{}, computed
with an aLIGO PSD and a uniform average over the sphere.

Relative to the $10\,\mathrm{ms}$ single-threaded CPU timing for
\textsc{lalsimulation}, the JAX surrogate is $10\times$ faster on an
L40S GPU for one waveform and ${\sim}140\times$ faster per waveform at
batch size~64. On CPU, JAX takes ${\sim}14\,\mathrm{ms}$ for one
waveform and ${\sim}1.9\,\mathrm{ms}$ per waveform at batch size~64
(Sec.~\ref{sec:speed}, Fig.~\ref{fig:latency}).

These speedups come from evaluating the networks for all 25 data pieces
together rather than sequentially (Sec.~\ref{sec:bank}) and expressing
waveform assembly entirely as tensor operations
(Sec.~\ref{sec:assembly}).
Phase integration is a precomputed matrix--vector product, the
Wigner-$D$ construction is a batched contraction, and strain projection
is a single Einstein summation over the modes. The full pipeline from
parameters to strain is compiled with \texttt{jax.jit}, while
\texttt{jax.vmap} provides parallelism across the batch.

Whittall \& Pratten~\cite{Whittall:2026yvp} likewise demonstrate
efficient GPU batching for a neural surrogate of a precessing waveform
model. Differences in the parent models, waveform duration, sampling,
hardware, and evaluation procedures prevent a quantitative comparison
of either accuracy or speed.
One qualitative difference in the tails is nevertheless worth noting:
their reported mismatch distribution extends to $\MM \gtrsim 10^{-1}$
(their Fig.~7), whereas at $M = 60\,\Msun$ our largest sky-averaged
mismatch over the $10\,000$ evaluation waveforms is $2.3\times10^{-3}$,
with a 99th percentile of $6.2\times10^{-4}$.  A difference of this size
in the tail is unlikely to be accounted for by the differing sky- and
polarization-averaging conventions alone.

\subsection{Applications}

Fast and accurate waveform models are a prerequisite for the next generation of gravitational-wave inference. A large body of work reduces the cost of likelihood evaluation through reduced-order quadrature~\cite{Canizares:2014fya,Smith:2016qas}, relative binning and heterodyned likelihoods~\cite{Zackay:2018qdy,Cornish:2021lje,Leslie:2021ssu}, and multibanding~\cite{Vinciguerra:2017ngf,Morisaki:2021ngj}, while a complementary line of work accelerates waveform generation itself, increasingly on GPUs. Our surrogate contributes to the latter, with the added property that the entire waveform-to-likelihood pipeline is differentiable. We highlight several capabilities that this combination unlocks, building on the GPU-accelerated \textsc{bilby} likelihood described in Sec.~\ref{sec:pe}.

The \textsc{JAX} implementation makes the entire signal pipeline (from surrogate evaluation through the time-to-frequency-domain conversion, detector response projection, and inter-site time delays) automatically differentiable. Forward-mode automatic
differentiation computes waveform Jacobians
$\partial\tilde{h}(f)/\partial\theta$ without the step-size tuning required by
finite differences.
This supports stable Fisher matrix calculations, including derivatives with
respect to the coalescence time. Since the calculation is JIT-compiled, it can
be vectorized over parameter space on a GPU to produce large grids of Fisher
matrices for population or network studies.

More broadly, the differentiable likelihood (Sec.~\ref{sec:pe}) is directly compatible with GPU-accelerated gravitational-wave samplers. \textsc{blackjax-ns}~\citep{Prathaban:2025qgg} re-implements the \textsc{bilby}+\textsc{dynesty} acceptance-walk algorithm within the \textsc{blackjax} framework~\cite{Cabezas2024arXiv240210797C}, achieving $20$--$40\times$ wall-time speedups using \textsc{ripple} waveforms~\cite{Edwards:2023sak}. The model presented here would extend this to generic precessing systems. Likewise, gradient-based samplers such as \textsc{flowMC}~\cite{Wong:2022xvh}, which combines NUTS and MALA kernels with normalizing-flow proposals, can exploit the exact likelihood gradients through the \textsc{Jim} interface \cite{Wong:2023lgb}.

The batched, JIT-compiled likelihood also makes importance sampling inexpensive. A large set of proposal samples from a \textsc{bilby}+\textsc{dynesty} run (including the typically $10$--$50\times$ more numerous weighted nested samples) can be re-evaluated on a GPU in seconds, for example to refine evidence estimates, enlarge the effective posterior, or correct a faster approximate likelihood back to the full one. For sampling-based inference the reweighting target is the same surrogate likelihood, so importance sampling is an optional refinement rather than a prerequisite. It becomes essential in simulation-based inference, where reweighting an approximate neural posterior against the exact likelihood is what renders the result asymptotically exact.
The same rapid waveform generation can accelerate simulation-based inference frameworks such as \textsc{Dingo}~\cite{Dax:2021tsq,Dax:2021myb,Dax:2022pxd,Dax:2024mcn}, both for producing large training sets of waveforms with faithful precession physics and for importance-sampling the resulting neural posterior estimates against the full likelihood.

Importance sampling \emph{across} waveform models, using the surrogate as a cheap proposal for the NRSur7dq4 likelihood to correct waveform systematics, is by contrast limited here by a smooth, parameter-dependent reference-phase offset between the two models. This reflects cross-model alignment rather than the surrogate's fidelity, does not affect inference performed self-consistently with either model, and we leave modeling the offset to enable efficient cross-model reweighting to future work.

Independently of inference speed, the surrogate natively accepts spins specified at a reference frequency (Sec.~\ref{sec:fref}), the standard convention in gravitational-wave parameter estimation, removing a conversion step that downstream pipelines would otherwise have to supply.
Together, these capabilities make the surrogate a practical engine for GPU-accelerated and gradient-based inference of precessing binaries.

\subsection{Limitations}

The NN surrogate inherits the calibration region of \nrsur{}, with mass
ratios up to 4 and spin magnitudes up to 0.8. \nrsur{} can be
extrapolated beyond this region. Although detailed NR simulations are
not available to test its accuracy thoroughly there, existing evidence
suggests that the extrapolation can remain reasonable when extending
either mass ratio or spin magnitude, but not both simultaneously. In
contrast, the NN surrogate
extrapolates poorly. This difference reflects the behavior of NN
activation functions outside the training distribution~\cite{xu2021how},
rather than the low-degree polynomial fits used in \nrsur{}.
Possible ways to improve this behavior include augmenting the training
set with extrapolated \nrsur{} waveforms, when such extrapolation is
considered acceptable, or modeling a residual relative to a PN
baseline.

The model presented here is deterministic and does not quantify its own
modeling error. A growing body of work constructs probabilistic waveform
models that describe distributions over waveforms rather than single
predictions~\cite{Moore:2014pda,Moore:2015sza,Doctor:2017csx,Williams:2019vub,Read:2023hkv,Khan:2024whs,Pompili:2024yec,Bachhar:2024olc}.
Such models can be incorporated into parameter estimation by marginalizing
over the additional degrees of freedom that describe waveform uncertainty,
thereby propagating modeling error into the posterior. Extending the
present architecture in this direction, for example by predicting
distributions for the data piece targets rather than point values, is a
natural direction for future work.

Finally, the post-merger smoothing of the orbital-frequency target
introduces hyperparameters.
At the level of the target waveforms, their effect is negligible. The
systematic change introduced by the conditioning is at least two orders
of magnitude below the surrogate's emulation error. This change is
included in all reported mismatches, which are computed against the raw
\nrsur{} waveforms, and is insensitive to the window choice
(Sec.~\ref{sec:smoothing}). We have not characterized how the
conditioning choice affects trainability or whether a different scheme
would allow the $\Omega$ networks to reach still lower error. A
systematic study would require regenerating the smoothed training data,
retraining the production-scale $\Omega$ networks, and repeating the
full mismatch evaluation for each configuration. Given the strong
dependence of accuracy on target quality (Sec.~\ref{sec:smoothing}), such
a study may yield further gains, but we leave it to future work.

\subsection{Broader impact and future directions}

The methodology developed here is not specific to \nrsur{}. The
piecewise decomposition, the emphasis on target quality, and the
error-budget-driven sizing provide a general approach to building fast,
differentiable surrogates of time-domain waveform models. We expect
these ideas to carry over to other members of the NRSurrogate family
and to effective-one-body models such as \textsc{SEOBNRv5PHM}, as well
as higher-dimensional problems involving remnant properties and
orbital eccentricity.

A related question is how direct per-node prediction scales to
longer waveforms, such as future NRSurrogate iterations or hybridized
models. \nrsur{} itself spans a fixed ${\sim}4400\,M$ ($\approx 20$
orbital cycles). This duration reaches the standard
$f_\mathrm{low}=20\,$Hz analysis band only for total masses
$M \gtrsim 55\,\Msun$. Lighter systems would require an extended
surrogate. At leading (0PN) order, the number of orbits from
$f_\mathrm{low}$ to merger depends only on the signal duration and not
on the masses,
$N_\mathrm{orb} = 0.8\,f_\mathrm{low}\,\tau$.
For equal-mass signals at $f_\mathrm{low}=20\,$Hz, durations of
$8$--$64\,$s correspond to $N_\mathrm{orb}\approx128$--$1024$ and
dimensionless durations of ${\sim}10^5$--$10^6\,M$.

The computational scaling is encouraging. The PieceMLP cost grows only
through the output layer ($\propto T$), while the TCMLP cost is linear
in $T$. The co-orbital-frame quantities remain slowly varying
regardless of duration. The number of time nodes~$T$ grows sublinearly
with duration. A time grid with a fixed number of nodes per orbit at
\nrsur{}'s density (${\sim}13$ nodes/orbit) yields a conservative upper
bound of ${\sim}1700$--$13000$ nodes over the $8$--$64\,$s range.
Because $\Omega(t)$ and the mode amplitudes are smooth in the
co-orbital frame, the required spline resolution is set by the
near-merger curvature, and the early inspiral should require far fewer
points.

The more restrictive requirement is the \emph{fractional} accuracy of
the orbital frequency. A coherent relative error
$\varepsilon = \delta\Omega/\Omega$ does not average to zero over the
inspiral and accumulates into an orbital-phase error
$\delta\phi = \int \varepsilon\,\Omega\,\dd t \sim 2\pi
N_\mathrm{orb}\,\bar\varepsilon$. Holding the end-to-end mismatch fixed
requires $\bar\varepsilon$ to decrease as $1/N_\mathrm{orb}$. Relative
to the present model, the required reduction is a factor of ${\sim}7$
at $8\,$s and ${\sim}60$ at $64\,$s, independently of how many time
nodes represent $\Omega(t)$. For long inspirals, regressing the
frequency residual against an analytic post-Newtonian or EOB baseline
would remove most of the accumulated phase while leaving the
merger-ringdown treatment unchanged.

\begin{acknowledgments}

The authors would like to thank Geraint Pratten, Lorenzo Pompili, and Stephen Green for helpful discussions and suggestions.
The authors acknowledge support from the URI Institute for AI \& Computational Research.
In particular, the computations were performed on the UMass-URI UNITY high-performance
computing (HPC) cluster hosted at the Massachusetts Green HPC Center.
M.P. is supported by NSF Grants AST-2407453 and PHY-2512902.
S.E.F.~acknowledges support from NSF Grants AST-2407454.
This research has made use of data or software obtained from the Gravitational Wave Open Science Center (gwosc.org), a service of the LIGO Scientific Collaboration, the Virgo Collaboration, and KAGRA. 
This material is based upon work supported by NSF's LIGO Laboratory which is a major facility fully funded by the National Science Foundation.
\end{acknowledgments}

\appendix

\section{Architecture sweep details}
\label{app:sweeps}

This appendix summarizes the alternative neural network architectures
we evaluated. Each architecture embodies different structural assumptions
about the function to be learned. We briefly describe the key ideas behind
each approach, why it might be suited (or unsuited) to waveform modeling,
and its empirical performance relative to our chosen TCMLP+SC architecture.

\begin{figure}[t]
  \centering
  \includegraphics[width=\columnwidth]{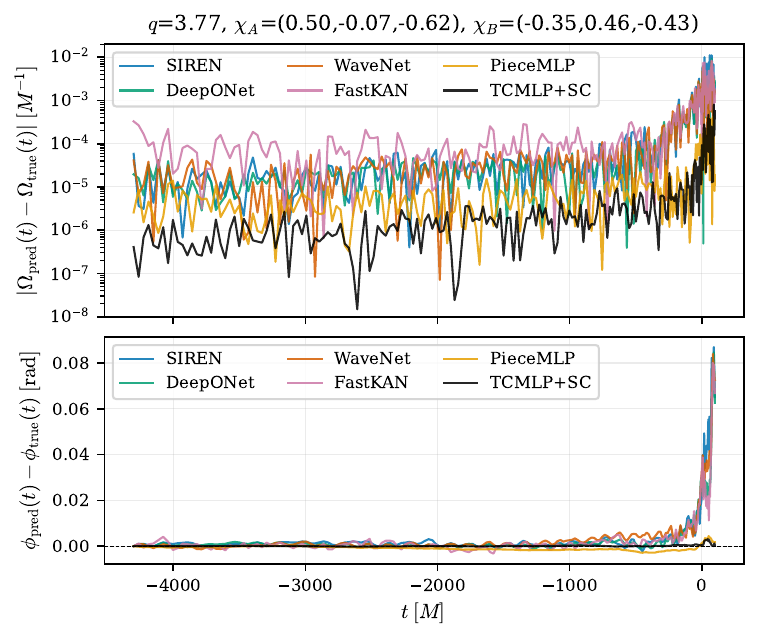}
  \caption{Architecture comparison on a representative validation waveform
    ($q = 3.77$, precessing spins).
    The top panel shows the absolute orbital-frequency residual
    $|\Omega_{\mathrm{pred}}(t) - \Omega_{\mathrm{true}}(t)|$ on a
    logarithmic scale.
    TCMLP+SC and PieceMLP maintain residuals ${\sim}10^{-6}$--$10^{-5}\,M^{-1}$
    throughout inspiral and merger, while WaveNet, DeepONet, SIREN, and FastKAN
    show $10$--$100\times$ larger errors near merger.
    The bottom panel shows the accumulated orbital-phase difference.
    The piecewise models accumulate ${\lesssim}5\,\mathrm{mrad}$ total
    phase error. The alternatives reach ${\sim}60\,\mathrm{mrad}$.}
  \label{fig:arch_comparison}
\end{figure}

We systematically evaluated several alternative neural-network
architectures for mapping a $7$-dimensional parameter vector to the
${\sim}230$-point non-uniform sparse time grid internal to NRSur7dq4.
To provide a fair comparison under production conditions, we retrained
WaveNet~\cite{Oord2016}, SIREN~\cite{Sitzmann2020,Mehta2021},
DeepONet~\cite{Lu2021deeponet}, and a FastKAN Kolmogorov-Arnold
Network~\cite{Liu2024kan,Li2024fastkan} on the same smoothed target
$\Omega_\mathrm{flt}(t)$
target and 5\,M training samples used for the production PieceMLP and
TCMLP+SC models, matching the parameter budget at $2.1$--$2.3$\,M
parameters each.
All architectures predict raw $\Omega(t)$, ensuring the same regression task.
Table~\ref{tab:arch_compare} reports the dimensionless validation MSE
(standardized by per-timestep variance of the training target),
and Fig.~\ref{fig:arch_comparison} shows a representative comparison
for a single validation waveform.

The piecewise models (PieceMLP and TCMLP+SC) outperform all four
alternatives by nearly two orders of magnitude in dimensionless MSE.
The advantage is even more striking for the accumulated orbital phase
error. PieceMLP and TCMLP+SC accumulate $\lesssim 5\,\mathrm{mrad}$
of phase error over the full waveform, whereas WaveNet, DeepONet,
SIREN, and FastKAN each accumulate $\sim 60\,\mathrm{mrad}$, a
${\sim}15\times$ larger error that would propagate into the assembled
gravitational waveform.
The FastKAN is the weakest of the alternatives tested, with a
dimensionless MSE comparable to SIREN and roughly $3\times$ above
WaveNet and DeepONet.

\begin{table}[b]
  \centering
  \caption{Architecture comparison on the orbital-frequency data piece
    (5\,M training samples, smoothed $\Omega$ target, 230-point sparse grid).
    All models are evaluated on the same 100\,K validation set.
    \emph{Dimless val MSE} is the per-timestep-variance--standardized MSE.
    PieceMLP and TCMLP+SC predict per-timestep outputs.
    The remaining architectures predict all 230 time steps simultaneously.}
  \label{tab:arch_compare}
  \begin{ruledtabular}
  \begin{tabular}{lrr}
    Architecture & Params & Dimless val MSE \\
    \hline
    \rule{0pt}{2.2ex}%
    TCMLP\,+\,SC ($384\times16$)  & 2.24\,M & $\mathbf{2.55\times10^{-5}}$ \\
    PieceMLP ($512\times12$)       & 3.01\,M & $4.06\times10^{-5}$ \\
    \hline
    DeepONet ($p{=}256$, $448\times6$) & 2.25\,M & $2.38\times10^{-3}$ \\
    WaveNet ($R{=}S{=}76$, $C{=}152$)  & 2.27\,M & $2.43\times10^{-3}$ \\
    SIREN ($384\times7$)               & 2.14\,M & $6.60\times10^{-3}$ \\
    FastKAN ($208\times5$, $G{=}12$)   & 2.33\,M & $7.05\times10^{-3}$ \\
  \end{tabular}
  \end{ruledtabular}
\end{table}

As a post-hoc diagnostic for the architecture comparison above, we
computed the empirical Neural Tangent Kernel (NTK) of each network
on a shared probe set of $N{=}8$ validation waveforms. The NTK,
$K_{ij}=\langle\partial_{\theta}f_i,\partial_{\theta}f_j\rangle$,
is the Gram matrix of per-output parameter gradients. Its
eigenspectrum governs which output-space directions gradient descent
represents most rapidly, and its decay rate is commonly used as a
proxy for a network's \emph{spectral bias}~\cite{jacot2018ntk,
pmlr-v97-rahaman19a}. The specific question motivating the check was
whether KAN-style parameterizations exhibit the broader at-init
eigenbasis recently claimed for shallow KANs~\cite{wang2024kan}, and
whether that property would predict final validation MSE on the
$\Omega(\vec\lambda,t)$ task. 
At initialization the six architectures produce qualitatively different
spectra (near-rank-one for SIREN/WaveNet/DeepONet vs. broad decays for FastKAN/TCMLP+SC/PieceMLP), but these differences neither track the
final MSE ranking nor persist through training. We include this
observation mainly as a null result.

\subsection*{WaveNet (causal dilated convolution)}

WaveNet-style architectures~\cite{Oord2016} process sequential data
by progressively combining information across increasingly large time
separations, building from local structure to global context through a
hierarchy of convolutional layers.
They use a stack of residual blocks with exponentially increasing
dilation rates $(1, 2, 4, \ldots, 512)$.
A dilation rate of $d$ means the convolution combines values separated
by $d$ time steps, allowing the network to aggregate information across
progressively larger time separations without increasing the number of
parameters. With this dilation pattern, the network's ``receptive field''
(the span of input points that influence each output) covers the full
230-point sequence.

The 7-dimensional parameter vector is projected to a per-channel
bias term that is broadcast over the time axis and added to each residual
block input. The architecture width is controlled by internal channel
counts $R{=}S{=}76$ and $C{=}152$ (see Ref.~\cite{Oord2016} for details).

The inductive bias~\footnote{
  An \emph{inductive bias} is a structural assumption built into a model
  that shapes what functions it can easily learn, independent of the training data.
}
of dilated convolutions is designed for
\emph{uniformly sampled} sequences, where a dilation of $d$ corresponds
to a physically meaningful time-shift $d \Delta t$.
On NRSur7dq4's non-uniform sparse grid, successive points span very
different physical durations (dense near merger, coarse during inspiral),
so the multi-scale structure encoded by the dilation pattern has no direct
physical interpretation.
At matched parameter count ($2.27\,\mathrm{M}$), WaveNet achieves a
dimensionless validation MSE of $2.75 \times 10^{-3}$, nearly
$108\times$ worse than TCMLP+SC (Table~\ref{tab:arch_compare}), while also
being substantially slower due to the sequential convolutional passes.

\subsection*{Modulated SIREN}

Sinusoidal Representation Networks~\cite{Sitzmann2020} (SIRENs) represent
a waveform as a continuous function of time, using sinusoidal internal
transformations that are naturally suited to representing smooth, oscillatory
signals, such as gravitational waveforms. \emph{Modulated} SIRENs~\cite{Mehta2021} 
add binary parameters controlling the shape of this function through learned
modulation coefficients.
SIRENs replace standard GELU/ReLU activations with $\sin(\omega_0 \cdot)$,
producing networks that are smooth functions of their inputs and that represent
high-frequency content more efficiently than piecewise-linear activations.

We implemented a \emph{modulated} SIREN~\cite{Mehta2021} in which the base
network takes the time coordinate as input and maps it to the output
via sinusoidal layers, while a small \emph{modulation network} maps
the 7D parameter vector to per-layer affine shifts and scales, so that
the time-series shape adapts to the binary parameters.
The variant reported in Table~\ref{tab:arch_compare} uses a base
network of $7$ hidden layers with $384$ neurons per layer
(denoted $384\times7$).
At $2.14\,\mathrm{M}$ parameters and 5\,M training samples on the
smoothed $\Omega$ target, SIREN achieves a dimensionless validation MSE
of $6.60 \times 10^{-3}$, $\sim 260\times$ worse than TCMLP+SC
(Table~\ref{tab:arch_compare}).
The phase residuals in Fig.~\ref{fig:arch_comparison} show that SIREN
exhibits large oscillatory errors near merger that integrate into
$\sim 60\,\mathrm{mrad}$ of accumulated phase error.

The key practical drawback is inference cost. Because SIREN represents
a continuous function $f(t;\theta)$ that must be evaluated separately
at each of the 230 grid points (rather than producing all 230 outputs
in a single pass), inference time scales with the number of output points,
raising GPU latency by ${\sim}66\times$ compared with a plain MLP.
This overhead, combined with the accuracy deficit, ruled SIREN out
for the latency-constrained bank-evaluation use case.

\subsection*{DeepONet}

Deep Operator Networks~\cite{Lu2021deeponet} (DeepONet) represent the
output as a weighted sum of learned basis functions (conceptually similar
to the reduced-basis or singular-value decompositions used in traditional
surrogate models) where the basis functions depend only on time and the
expansion coefficients depend only on the binary parameters.

Concretely, a DeepONet decomposes the output function as a learned inner
product between a \emph{branch network} (7D parameters $\to$ $p$ basis
coefficients) and a \emph{trunk network} (scalar time $t$ $\to$ $p$ basis
functions).
\begin{equation}
  \hat{\Omega}(t;\,\bm{\lambda}) \;=\;
  \sum_{k=1}^{p} c_k(\bm{\lambda})\, b_k(t) + q_0 \,.
\end{equation}
This is the most natural operator-learning representation for our task
and, in theory, confers the ability to evaluate at arbitrary time
coordinates without retraining.
The variant reported in Table~\ref{tab:arch_compare} uses basis rank
$p=256$ with branch and trunk networks both of $6$ hidden layers
with $448$ neurons per layer (denoted $p{=}256$, $448\times6$).

In practice, the assumption that the output can be written as a sum of
separable terms which are each a product of a parameter-dependent coefficient
and a time-dependent basis function proved too restrictive.
The orbital frequency near merger depends on the binary parameters and
time in a strongly non-separable way (the merger time itself is a
function of $\bm{\lambda}$), and the rank-$p$ factorization cannot
represent this with a fixed number of modes.
At $2.25\,\mathrm{M}$ parameters and 5\,M training samples,
DeepONet achieves a dimensionless validation MSE of
$2.38 \times 10^{-3}$, $\sim 90\times$ worse than TCMLP+SC
(Table~\ref{tab:arch_compare}), and accumulates $\sim 60\,\mathrm{mrad}$
of orbital phase error (Fig.~\ref{fig:arch_comparison}).

\subsection*{Kolmogorov-Arnold Networks}

Kolmogorov-Arnold Networks~\cite{Liu2024kan} (KANs) generalize
standard neural networks by replacing the fixed nonlinear activation
functions with flexible, learnable univariate functions, aiming to
represent complex mappings more efficiently by adapting the nonlinearities
to the problem at hand.
This is motivated by the Kolmogorov-Arnold representation
theorem~\cite{Kolmogorov1956Representation}, which guarantees that any multivariate
continuous function can be written as compositions of univariate functions.

The original KAN formulation uses cubic B-splines to represent these
learnable functions and has been claimed to achieve 10--100$\times$
parameter efficiency over MLPs on smooth, low-dimensional scientific targets.
However, evaluating B-splines requires conditional operations (finding
which spline segment is active) that are inefficient on GPUs, making
these networks 10–100$\times$ slower per parameter than standard architectures.

We therefore tested FastKAN~\cite{Li2024fastkan}, which replaces each B-spline
with a sum of Gaussian basis functions centered on a fixed grid. This approximation
can be evaluated efficiently with standard tensor operations. At matched parameter
count, FastKAN trains and evaluates at speeds comparable to standard networks.

We trained FastKAN with $2.33\,\mathrm{M}$ parameters with architecture
$[7, 208, 208, 208, 208, 208, 230]$ (five hidden layers of width $208$, with $G=12$ Gaussian
basis functions per connection) on the same data and training protocol as the other
architectures. FastKAN achieves a dimensionless validation MSE of
$7.05 \times 10^{-3}$, making it the weakest architecture tested,
${\sim}276\times$ worse than TCMLP+SC. This is consistent with
Ref.~\cite{Shukla_2024CMAME.43117290S}, which found that KAN-style models do
not outperform standard networks on smooth regression tasks when capacity and
training budgets are matched fairly. We attribute part of the gap to our
7-dimensional input space. Reported advantages for KANs are strongest in
2–5 dimensions, and the benefit appears to dilute at higher dimensionality. 
While FastKAN handles vector outputs straightforwardly by setting the final layer width to the number of output coordinates, 
the Kolmogorov–Arnold theorem motivates scalar-valued representations and therefore applies to our target only componentwise. 
In practice, the model must learn separate output-coordinate edge functions or basis combinations for all 230 time steps, 
which may dilute the parameter-efficiency advantage seen in smaller scalar-output settings.
Additionally, the smoothness in time that motivates KANs here is already captured
by our Time-Conditioned MLP through Fourier feature encoding of the time coordinate.

\subsection*{TCMLP hyperparameter study}
\label{app:tcmlp_ablations}

The Time-Conditioned MLP (TCMLP) introduced in Sec.~\ref{sec:tcmlp}
requires three non-obvious hyperparameter choices. These are the Fourier
feature encoding dimension, residual connections, and batch size. Because
these choices significantly affect accuracy and training stability, and
because optimal values may differ from common defaults in the ML literature,
we report a systematic hyperparameter (ablation) study here.
Table~\ref{tab:tcmlp_ablation} summarizes results at 2\,M training samples.

\begin{table}[t]
  \centering
  \caption{TCMLP hyperparameter study on the orbital-frequency data piece
    (2\,M samples, standardized MSE target, 300 epochs unless noted).
    The \texttt{PieceMLP h512\_d12} baseline has
    MSE$\,=3.85\times10^{-6}$.
    Entries marked $\dagger$ were cancelled at wall-clock limit and
    were still improving.}
  \label{tab:tcmlp_ablation}
  \begin{ruledtabular}
  \begin{tabular}{lrrrr}
    Config & $H$ & $D$ & FFE & Val MSE \\
    \hline
    h256\_d16\_ffe16\_res           & 256 & 16 & 16 & $5.93\times10^{-6}$ \\
    h256\_d20\_ffe16\_res           & 256 & 20 & 16 & $\mathbf{5.31\times10^{-6}}$ \\
    h256\_d24\_ffe16\_res$^\dagger$ & 256 & 24 & 16 & ${\sim}5.52\times10^{-6}$ \\
    h256\_d16\_ffe32\_res           & 256 & 16 & 32 & $6.34\times10^{-6}$ \\
    h256\_d16\_ffe16\_res (bs=512)  & 256 & 16 & 16 & $7.05\times10^{-6}$ \\
    h384\_d16\_ffe16\_res$^\dagger$ & 384 & 16 & 16 & ${\sim}5.85\times10^{-6}$ \\
    h512\_d12\_ffe16\_res$^\dagger$ & 512 & 12 & 16 & ${\sim}6.89\times10^{-6}$ \\
  \end{tabular}
  \end{ruledtabular}
\end{table}

The ablation reveals several patterns.
\begin{itemize}
  \item \emph{Depth helps over the range probed.}  Increasing from
    $D=16$ to $D=20$ reduces MSE by ${\sim}10\%$.  The $D=24$ run was
    cancelled before convergence and appears close to $D=20$, so we
    cannot say where the trend turns over.
  \item \emph{More FFE frequencies did not help.}  FFE-32 increases MSE
    by $7\%$ relative to FFE-16 at identical depth and width, on the
    single pair of runs tested.  Sixteen Fourier frequencies appear
    sufficient for the range of timescales in the 230-point grid.
  \item \emph{Batch size matters.}  Doubling from $256$ to $512$ at
    identical architecture degrades MSE by ${\sim}19\%$; we did not scan
    the learning rate jointly, so part of this may be recoverable.
  \item \emph{Residual connections are essential for $D\geq12$.}
    Without residuals, depth-12 TCMLP training diverges.  With
    residuals, training remains stable up to $D=24$.
  \item \emph{The unconditioned TCMLP trails PieceMLP in this sweep.}
    The best fully converged TCMLP of this sweep
    (h256\_d20\_ffe16\_res, MSE$\,=5.31\times10^{-6}$), which uses
    neither spin conditioning nor a wider layer, is $1.38\times$ worse
    than PieceMLP (\texttt{h512\_d12}, $3.85\times10^{-6}$) at 2\,M
    samples.  Adding spin conditioning closes the gap already at this
    sample size: TCMLP+SC (\texttt{h384\_d16}) reaches
    $3.83\times10^{-6}$, marginally ahead of PieceMLP
    (Table~\ref{tab:tcmlp_vs_piecemlp}).  At 5\,M samples with the
    smoothed $\Omega$ target the production TCMLP+SC model reaches
    $2.55\times10^{-5}$ against $4.06\times10^{-5}$ for PieceMLP, a
    $37\%$ improvement; the training set, the target and the spin
    conditioning all differ between these two comparisons
    (Sec.~\ref{sec:tcmlp}).
\end{itemize}

Taken together, these results indicate that TCMLP benefits from moderate depth
($D \approx 20$) with residual connections, a Fourier feature encoding matched to
the timescales present in the data (FFE-16 suffices), and relatively small batch
sizes that provide noisier but more frequent gradient updates. The architecture
is not parameter-efficient at small training set sizes, but this gap closes—and
reverses—as the training set grows, consistent with the general observation that
simpler, more flexible architectures benefit disproportionately from additional data.

\section{Prior preservation under the reference-frequency spin map}
\label{app:prior_map}

In the parameter-estimation application (Sec.~\ref{sec:pe}) we sample with
the spin vectors defined at the neural surrogate start time $t_0$ and map the
resulting posterior to the conventional reference-frequency convention in a
single post-processing pass (Sec.~\ref{sec:fref}).  Here we show that this
procedure is equivalent to sampling directly at $f_\mathrm{ref}$ exactly at
the level of the likelihood; that the spin magnitudes and the global
orientation content are preserved exactly; and that the residual prior
difference, which affects the relative spin-tilt sector and the overall
prior volume, is small and precession-dependent.

Let $\lambda=(m_1,m_2,\bm{\chi}_1^{(t_0)},\bm{\chi}_2^{(t_0)},\xi)$ denote the
parameters in the $t_0$ convention, with $\xi$ the extrinsic parameters, and
let $\mu=T(\lambda)$ be the corresponding parameters at $f_\mathrm{ref}$,
where $T$ is the bijective spin-evolution map of Sec.~\ref{sec:fref}
($\lambda\equiv\theta_{t_0}$ and $\mu\equiv\theta_{f_\mathrm{ref}}$ in the
notation there).
The two conventions are arranged to emit the \emph{same} physical
waveform, $h_{t_0}(\lambda)=h_{f_\mathrm{ref}}(T(\lambda))$, and the
matched-filter likelihood is a functional of that waveform, so
$L_{t_0}(d\mid\lambda)=L_{f_\mathrm{ref}}(d\mid T(\lambda))$.  The
pushforward of the $t_0$ posterior under $T$ therefore equals the posterior
obtained by sampling at $f_\mathrm{ref}$ if and only if the priors satisfy
the pull-back relation
\begin{equation}
  \pi_{t_0}(\lambda) \;=\; \pi_{f_\mathrm{ref}}\!\big(T(\lambda)\big)\,
  \big|\det J_T(\lambda)\big| ,
  \label{eq:pullback}
\end{equation}
with $J_T=\partial\mu/\partial\lambda$.  The prior is the only place the two
analyses can differ.

The map $T$ fixes the masses and extrinsic parameters and acts on the two
Cartesian spin vectors, $\bm{\chi}^{(f_\mathrm{ref})}=R(\lambda)\,
\bm{\chi}^{(t_0)}$ with $R\in SO(3)\times SO(3)$.  Ordering the parameters
as $(m,s,\xi)$ with $s$ the six Cartesian spin components, the Jacobian is
block-triangular,
\begin{equation}
  J_T =
  \begin{pmatrix}
    I_m & 0 & 0 \\[2pt]
    F_{,m} & F_{,s} & 0 \\[2pt]
    0 & 0 & I_\xi
  \end{pmatrix},
  \qquad
  \det J_T = \det F_{,s} ,
  \label{eq:jacobian}
\end{equation}
so only the $6\times6$ spin block $F_{,s}$ contributes.

It is tempting to identify $F_{,s}$ with $R$ and conclude that
$\det J_T=1$, but that does not hold here.  First, $R$ depends on the spin
configuration itself, so
$(F_{,s})_{ab}=\delta_{ab}R_a+C_{ab}$ with $(C_{ab})_{ij}=\sum_k
\partial (R_a)_{ik}/\partial (s_b)_j\,(s_a)_k$, including cross-spin blocks
$C_{12}$ and $C_{21}$.  Second, the reference time $\tau$ is defined
implicitly by $\Omega(\tau;\lambda)=\omega_\mathrm{ref}$, so the map is
evaluated on a parameter-dependent surface; writing the stacked spin
trajectories as $X(t;\lambda)$, implicit differentiation gives
\begin{equation}
  F_{,s} = X_{,s} - \frac{X_{,t}\,\Omega_{,s}}{\Omega_{,t}} ,
  \label{eq:Fs}
\end{equation}
a rank-one correction to the fixed-time flow.

Evaluating $\det J_T$ by finite differences of the forward map at
$f_\mathrm{ref}=20\,\mathrm{Hz}$, for draws spanning the calibration
region, gives values between $0.89$ and $1.03$: departures from unity of a
few percent, with a median of ${\sim}3\%$ and a largest value of
${\sim}11\%$.  Holding $\tau$ fixed gives a comparable spread, so the
effect is dominated by the cross-spin terms in $X_{,s}$ rather than by the
moving endpoint.  These values use the neural spin trajectories and so
include the spin networks' own error; they are best read as an upper bound
on the distortion of the exact map.

The prior is therefore reshaped at the percent level rather than preserved
exactly, and Eq.~\eqref{eq:pullback} does not reduce to
$\pi_{t_0}=\pi_{f_\mathrm{ref}}\circ T$.  What remains exact is more
limited: $T$ preserves each spin magnitude $|\bm{\chi}_i|$ identically, and
it is equivariant under a global rotation, so a globally isotropic
orientation prior remains globally isotropic.  The canonical
``isotropic-direction, uniform-magnitude'' prior is invariant in these two
senses, and the magnitude priors and $\chieff$ are correspondingly
unaffected.

Bilby~\cite{Ashton2019} samples the spin magnitudes and orientation angles
$(a_i, \theta_i, \phi_{12}, \phi_{JL}, \theta_{JN})$, defined in full in
Ref.~\cite{Romero-Shaw:2020owr}. The tilts $\theta_i$
are measured from the (Newtonian) orbital angular momentum $\hat{L}$ and,
with the in-plane angle $\phi_{12}$, constitute the relative-angle sector,
while $\phi_{JL}$ and $\theta_{JN}$ orient the orbital and total angular
momenta relative to the total angular momentum $J$ and the line of sight,
constituting the global-orientation sector.
The relative-angle sector is \emph{not} automatically preserved.  A single global
rotation leaves the isotropic prior invariant, but $R$ depends on the
configuration itself, so two-spin nutation reshapes the joint distribution
of the tilts and $\phi_{12}$ between $t_0$ and $f_\mathrm{ref}$.  Hence
$T$ preserves the magnitude and global-orientation content exactly, while
the spin-tilt sector is reshaped.

We bound the residual with a likelihood-free prior-predictive test. We draw
the canonical prior at $t_0$, forward-evolve it to
$f_\mathrm{ref}=20\,\mathrm{Hz}$ with the exact \nrsur{} spin ODE over the
calibration region ($q\in[1,4]$, $|\bm{\chi}_i|\le0.8$,
$M\in[60,120]\,\Msun$) for $10^4$ draws, and compare the $t_0$ and remapped
marginals with the Jensen--Shannon (JS) divergence~\cite{Romero-Shaw:2020owr}.
The $\cos\theta_i$ and $\phi_{12}$ marginals agree with their $t_0$
form to $\lesssim2\times10^{-3}\,\mathrm{bit}$. This is at or below both the
Monte-Carlo noise floor for $10^4$ samples (${\approx}1.8\times10^{-3}\,\mathrm{bit}$)
and the ${\sim}2\times10^{-3}\,\mathrm{bit}$ value conventionally taken as
indistinguishable.
These marginals also remain below the noise floor even when restricted to the
most strongly precessing draws ($\chip\in[0.5,0.8)$). The spin magnitudes and
$\chieff$ have one-dimensional marginals preserved to
${\sim}10^{-4}\,\mathrm{bit}$.  Individual samples
nonetheless move by $\langle|\Delta\cos\theta_i|\rangle$ up to ${\sim}0.07$
at high $\chip$, so the joint reshaping is physical but does not register at
the level of the marginal priors.

The $t_0$ posterior therefore pushes forward under $T$ to the
$f_\mathrm{ref}$ posterior up to a prior difference that is
indistinguishable at the level of the one-dimensional marginals and at
the present Monte-Carlo resolution.  We have not tested the joint
distribution, which the per-sample tilt shifts above show is genuinely
reshaped; the tests here bound the marginals only.  The difference is
negligible for the weakly precessing GW150914
(Fig.~\ref{fig:pe_gw150914}).
For strongly precessing systems an exact match to the $f_\mathrm{ref}$ prior
can be recovered by importance-reweighting the remapped samples by the prior
ratio.  We stress that this physical prior shift is distinct from (and much
smaller than) the error incurred by an \emph{inaccurate} remap. Propagating
the same draws through the neural network's fitted spin trajectory rather
than the exact ODE inflates the apparent tilt shift by an order of magnitude
and produces anomalously large changes in $\chieff$, violating its
expected approximate conservation~\cite{Racine:2008qv}, and so is not a
measure of the prior difference.

\bibliography{references.bib}

\end{document}